\documentclass[12pt]{article}
\usepackage{amsmath}
\usepackage{graphicx}
\usepackage{enumerate}
\usepackage{natbib}
\usepackage{amstext}
\usepackage{url}

\usepackage{amssymb}
\usepackage{subfigure}
\usepackage{color,xcolor}
\usepackage[left=1in,right=1in,top=1in,bottom=1in]{geometry}
\usepackage[colorlinks,linkcolor=blue,anchorcolor=blue,citecolor=blue,CJKbookmarks=True]{hyperref}

\newtheorem{theorem}{Theorem}

\newtheorem{corollary}{Corollary}
\newtheorem{proposition}{Proposition}

\renewcommand{\hat}{\widehat}

\newcommand{\bZ}{\mbox{\bf Z}}

\newcommand{\bX}{\mbox{\bf X}}

\newcommand{\by}{\mbox{\bf y}}
\newcommand{\bx}{\mbox{\bf x}}

\newcommand{\bH}{\mbox{\bf H}}
\newcommand{\bI}{\mbox{\bf I}}

\newcommand{\bD}{\mbox{\bf D}}

\newcommand{\bw}{\mbox{\bf w}}

\newcommand{\bG}{\mbox{\bf G}}

\newcommand{\bbeta}{\mbox{\boldmath${\beta}$}}

\newcommand{\bSigma}{\mbox{\boldmath${\Sigma}$}}

\newcommand{\bepsi}{\mbox{\boldmath${\varepsilon}$}}

\newcommand{\bPhi}{\mbox{\boldmath${\Phi}$}}

\newcommand{\bmu}{\mbox{\boldmath${\mu}$}}

\newcommand{\bpsi}{\mbox{\boldmath${\psi}$}}

\newcommand{\bOmega}{\mbox{\boldmath${\Omega}$}}

\def\tilde{\widetilde}

\def\diag{\mbox{diag}}

\begin{document}

\def\spacingset#1{\renewcommand{\baselinestretch}%
{#1}\small\normalsize} \spacingset{1}

%
%

  \title{\bf Optimal integrating learning for split questionnaire design type data}
  \author{Cunjie Lin$^{a,b}$, Jingfu Peng$^b$\thanks{The first two authors contributed equally.}, Yichen Qin$^c$, Yang Li$^{a,b}$\thanks{Correspondence to Yang Li (yang.li@ruc.edu.cn).}, and Yuhong Yang$^d$\\
{\footnotesize $^{a}$Center for Applied Statistics, Renmin University of China, Beijing, China;}\\
{\footnotesize $^{b}$School of Statistics, Renmin University of China, Beijing, China;}\\
{\footnotesize  $^c$Department of Operations, Business Analytics, and Information Systems, University of Cincinnati, OH, USA;}\\
{\footnotesize $^d$School of Statistics, University of Minnesota, MN, USA}}
  \maketitle

\bigskip
\begin{abstract}

In the era of data science, it is common to encounter data with different subsets of variables obtained for different cases. An example is the split questionnaire design (SQD), which is adopted to reduce respondent fatigue and improve response rates by assigning different subsets of the questionnaire to different sampled respondents. A general question then is how to estimate the regression function based on such block-wise observed data. Currently, this is often carried out with the aid of missing data methods, which may unfortunately suffer intensive computational cost, high variability, and possible large modeling biases in real applications. In this article, we develop a novel approach for estimating the regression function for SQD-type data. We first construct a list of candidate models using available data-blocks separately, and then combine the estimates properly to make an efficient use of all the information. We show the resulting averaged model is asymptotically optimal in the sense that the squared loss and risk are asymptotically equivalent to those of the best but infeasible averaged estimator. Both simulated examples and an application to the SQD dataset from the European Social Survey show the promise of the proposed method.
\end{abstract}

\noindent%
\vfill

\newpage
\spacingset{1.5} 
\section{Introduction}
\label{sec:intro}

Recent advances in data collection technologies make it common to collect data with different subsets of variables obtained for different cases. Examples can be found in financial technology, medical research as well as large scale survey. Take credit risk assessment for instance, credit rating agencies collect individual information, such as mobile payment history and online shopping records, to conduct an efficient credit risk assessment on internet loan applicants \citep{Fang:2017}. However, it might be infeasible to collect all items since some applicants may not have shopping records in certain platforms at all, and there is no need to conduct complete collection as it may either take more time, inflate administration cost, and induce break-off of application. Consequently, only a small fraction of individuals have all the information available while most individuals have only a subset of covariates observed in the data base. Such block-wise observed data are also common in clinical research where data with different measurements in several uncoordinated electronic medical record systems are pooled together to understand the mechanism of diseases \citep{Wang2019emr}.
This kind of data has also been intentionally collected by survey practitioners, which is also known as split questionnaire design (SQD) data \citep[][]{Graham:2006, Rhemtulla:2016Planned}. With only a subset of questions being asked for each respondent, respondents may become more engaged during the survey, hence resulting in higher response rates \citep[][]{Andreadis2020} and higher data quality \citep{Toepoel:2018}. Hereafter such block-wise data are referred to collectively as the ``SQD-type data'' in this study.

Given the increasing and potentially more pragmatic benefits of SQD-type data, it is important to develop efficient statistical methods capable of analyzing such data. Some progress has been made in the area of SQD for simple descriptive targets, such as the population mean \citep{RenssenandNieuwenbroek1997, Gelman:1998, Merkouris2004} with results on optimal questions split \citep{AdiguzelandWedel:2008, ChipperfieldandSteel:2009, Stuart:2019}. However, a major challenge for estimating regression function based on SQD-type data arises due to its high missing rate and block-wise observed data structure. Typically, the data collected have only a small proportion of cases with complete observations across all subsets. And each incomplete case has measurements only for a subset of variables and misses most of the information in remaining subsets.

One possible strategy is to treat it as a missing data problem and the simplest method is to perform complete cases (CC) analysis. However, CC method is inefficient since it fails to utilize a large amount of information in the incomplete cases.
Alternatively, full information maximum likelihood \citep{Allison2012, Chipperfield:2018} and multiple imputation (MI) \citep[][]{rubin2004multiple, AdiguzelandWedel:2008} are commonly used to deal with missing data, but poor performance is witnessed
when the distribution assumptions are violated and too many covariates are missing. In view of the high missing rate and large number of missing covariates, the traditional techniques for missing data have been rendered improper in analysis of SQD-type data.

Model averaging is an alternative approach to estimating regression function utilizing the SQD-type data. It works by building a list of candidate models using available information and aggregating them with proper weights. A growing body of literature has focused on this topic when data are fully observed, see \cite{Hoeting1999}, \cite{Yang2001}, \cite{Moral-Benito:2015} and the references therein.
For missing covariates data, there have also been some model averaging approaches. For example, a Mallows model averaging method is proposed in \cite{Zhang:2013} but only after replacing the missing values with zeros, which may result in biased estimates when the missing values are significantly nonzero. In a similar direction with \cite{Zhang:2013}, \cite{Fang:2017} proposed to select the optimal weight using delete-one cross-validation based on complete cases. Although it avoids imputing the missing values, the weights learning procedure utilizes only the complete cases but ignoring large proportion of incomplete cases, which may not be capable of exploiting all the benefits of the SQD-type data considered in this article. Further, both approaches focus on linear candidate models with least square estimation, which may not be flexible and efficient enough in real applications.

In this article, we propose a novel method called split-questionnaire averaged regression estimation (SQUARE). We first construct a list of candidate regression models based on different observed data-blocks, including a complete cases data-block and several incomplete cases data-blocks.
For data-block of the complete cases, the regression estimator has the possibly smallest bias but the largest variance due to the small simple size. In contrast, the regression models based on the incomplete cases data-blocks could result in a significant reduction in estimation variance since more cases are used but fewer parameters are estimated. However, each of them only captures certain aspects of the signal, hence leading to a possibly substantial bias. To integrate the different parts, a proper weight vector is determined
 to combine these candidates for the purpose of balancing the bias and variance. The proposed criterion incorporates information of both complete and incomplete cases and meanwhile measures the relative contributions of each candidate in predicting the responses of complete cases data.
 Different from most model averaging, in our context, we allow the model weights to vary freely between 0 and 1, so that the complementary signals in different subsets can be efficiently combined. The excellent properties of the proposed approach can be well verified theoretically and numerically.


The remainder of this article is organized as follows. In Section \ref{sec:model}, we introduce some notations about the SQD-type data explicitly and present the details of the proposed SQUARE procedure. Section \ref{sec:property} presents theoretical properties. Section \ref{sec:simulation} investigates finite sample performance of SQUARE by numerical examples. We apply our method to ESS dataset in Section \ref{sec:realdata}. Concluding remarks and discussions are in Section \ref{sec:discussion}. The proofs of the main results are in the \hyperref[appendix]{Appendix}.

\section{Methodology}
\label{sec:model}
\subsection{Notation}

Consider the data-generating process
\begin{equation}\label{eq:model}
\begin{split}
Y&=\mu(\bX)+\varepsilon,\\
\end{split}
\end{equation}
where $Y$ is the response variable, $\bX=(X_{1},\ldots,X_{p})^{T}$ is a $p$-dimensional covariate vector, $\varepsilon$ is a random error term with $E(\varepsilon)=0$ and finite variance $E(\varepsilon^2) < \infty$, and $\mu(\bx)$ is the conditional mean of $Y$ given $\bX=\bx$. Our goal is to estimate regression function $\mu$ based on the block-wise observed data.

 Suppose the response is always observed and the $p$-dimensional covariates are divided into $M+1$ distinct subsets $\Delta_0,\Delta_1,\cdots, \Delta_{M}$, where $\Delta_i \subset \{1,\cdots, p\}$, $i=0,1,\cdots, M$. Let $p_m=|\Delta_m|$, where $|\Delta|$ denotes the cardinality of a set $\Delta$ and $p=\sum_{m=0}^{M}p_m$. We consider the scenario of block-wise observed data, where only a small size of samples can be observed for all covariates $\bigcup_{m=0}^{M}\Delta_m$, while a large proportion of cases are missing for most subsets of covariates. Specifically, let $S_0$ be the index set of complete cases with all covariates observed and $S_m$ be the index set of cases with available subsets $\Delta_0$ and $\Delta_m$, $m=1,\cdots, M$. Here, $\Delta_0$ is the common set containing covariates available for all cases. It is possible that $\Delta_{0}$ contains no covariates other than an intercept term, or even is an empty set $\varnothing$. For $m=0,1,\cdots, M$, let $n_m=|S_m|$ denote the sample size of cases in $S_m$, in practice, $n_0$ would be much smaller than $n_m$, but it is still expected to be larger than $p$. For any $S_m$ and $\Delta_k$, let $\by_{S_m}$ and $\bepsi_{S_m}$ denote the responses and random error terms of cases in $S_m$ respectively. And let $\bX_{S_m\Delta_k}$ represent data matrix of cases in $S_m$ with variables in $\Delta_k$. In these notations, for $m\geq 1$ and $k\geq 1$, $\bX_{S_m\Delta_k}$ is missing when $m \neq k$.

\begin{figure}[!htbp]
\centering
\includegraphics [angle=-0, scale=0.9]{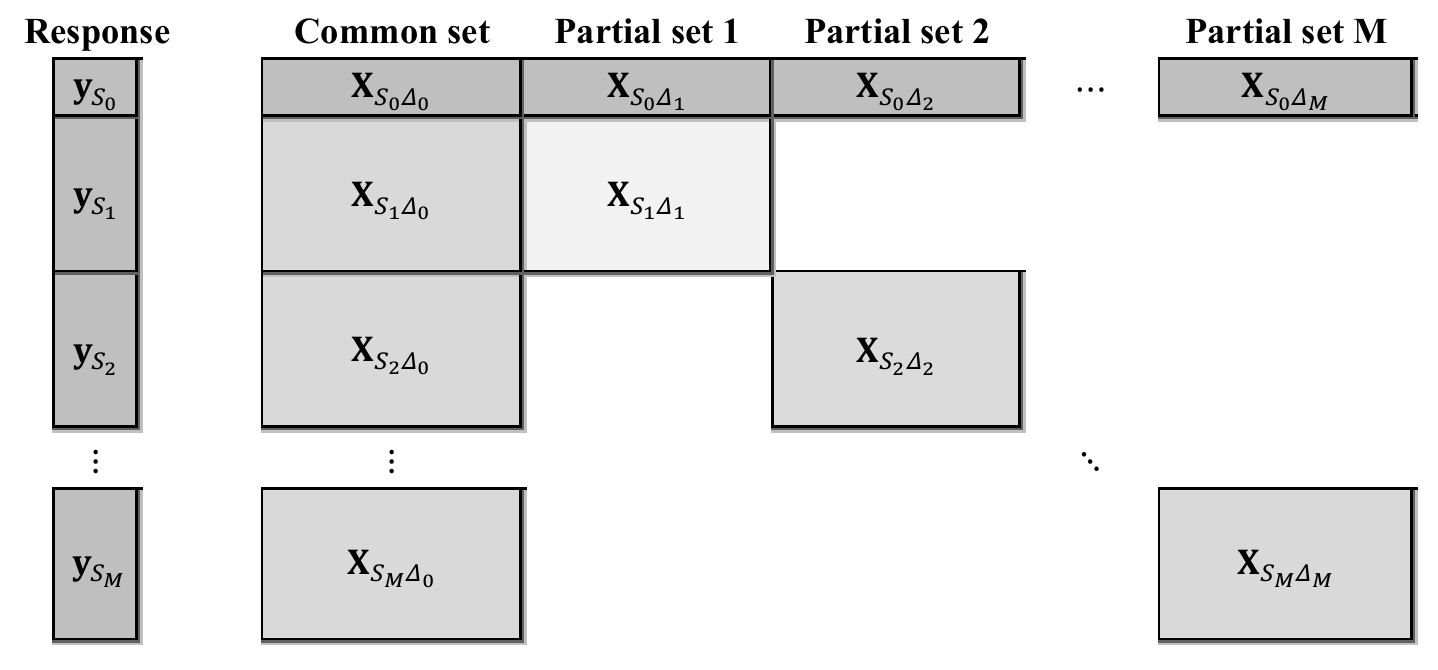}\par
\caption{The structure of SQD-type data. The grey blocks represent the observed data.}
\label{fig:structure}
\end{figure}

 By rearranging the order of the covariates and cases, we can obtain a simple pattern of the block-wise observed data as exhibited in Figure~\ref{fig:structure}, in which grey blocks represent observed data. Note that this block-wise observed data is commonly encountered in SQD study which splits long questionnaire into shorter parts and each respondent answers one of the sub-questionnaires and thus termed ``SQD-type data'' in this study. Nonetheless, the SQD-type data is also encountered in many modern scientific researches where data are collected from multiple modalities, or according to a planned missing design \citep{Graham:2006, Rhemtulla:2016Planned} as discussed in Section \ref{sec:intro}.


\subsection{Candidate Estimators}
To estimate the regression function based on the SQD-type data, we first divide the whole dataset into $M+2$ data-blocks $\{\bD_m\}_{m=0}^{M+1}$ according to the division of covariates, where $\bD_0=\left[\by_{S_0},\bX_{S_0,\bigcup_{m=0}^M\Delta_m}\right]$, $\bD_1=\left[\by_{\bigcup_{m=1}^M S_m},\bX_{\bigcup_{m=1}^M S_m,\Delta_0}\right]$, and $\bD_{m+1}=\left[\by_{S_m},\bX_{S_m\Delta_m}\right]$ for $m\geq 1$. Here, $\by_{\bigcup_{m=1}^MS_m}=(\by_{S_1}^T,\cdots,\by_{S_m}^T)^T$, $\bX_{S_0,\bigcup_{m=0}^M\Delta_m}=(\bX_{S_0\Delta_0},\cdots, \bX_{S_0\Delta_M})$, and
$\bX_{\bigcup_{m=1}^M S_m,\Delta_0}=(\bX_{S_1\Delta_0}^T,\cdots,\bX_{S_M\Delta_0}^T)^T$. Based on these $M+2$ data blocks, we can construct $M+2$ candidate estimators for the regression function $\mu$. In this study, the candidate models on the block observations are not restricted to linear regression, as long as the associate estimators are linear with respect to the response variable. Let $\hat{\mu}(\bx; \bD_m)$ be the candidate estimator based on $\bD_m$, the aggregated estimator of the regression function $\mu$ can be defined as
\begin{equation}\label{eq:estimatortrain}
  \hat{\mu}(\bw)=\sum_{m=0}^{M+1}w_m\hat{\mu}(\bx; \bD_m),
\end{equation}
where $\bw=(w_0,w_1,\ldots, w_{M+1})^{T}$  is a weight vector belonging to the set $Q=\{\bw \in[0,1]^{M+2} : 0 \leq w_{m} \leq 1\}$.

There are a variety of methods to construct regression estimators. Considering the smaller sample size of the complete cases, we focus on the linear regression model with $\bD_0$ and the candidate estimator can be obtained by least squares estimation, i.e,
\[
\hat{\mu}(\bx;\bD_0)=\bx^T\left(\bX_{S_0,\cup_{m=0}^M\Delta_m}^T\bX_{S_0,\cup_{m=0}^M\Delta_m}\right)^{-1}
\bX_{S_0,\cup_{m=0}^M\Delta_m}^T\by_{S_0}.
\]
For data blocks $\bD_m, m\geq 1$, we focus on the estimator which is linear with respect to the response variable. Specifically, $\hat{\mu}(\bx; \bD_1)=\bpsi_1^{T}(\bx; \bX_{\bigcup_{m=1}^M S_m,\Delta_0})\by_{\bigcup_{m=1}^M S_m}$ and $\hat{\mu}(\bx; \bD_{m+1})=\bpsi_{m+1}^{T}(\bx; \bX_{ S_m\Delta_m})\by_{S_m}$ for $m\geq 1$, where $\bpsi_1$ and $\bpsi_{m+1}$ are $(\sum_{m=1}^{M}n_m)$-dimensional and $n_m$-dimensional vector respectively that used to linearly combine the response variable.  Note that the popular estimators such as least squares, ridge regression, Nadaraya-Waston and local polynomial kernel regression with fixed bandwidths, nearest neighbor estimators, series estimators and spline estimators are all special cases of linear estimator. Here, we are concerned primarily with the linear estimator instead of linear regression models, since the linear estimator has extensively been applied in both linear regression and nonlinear regression models due to its satisfactory statistical properties and computational advantages.

\begin{figure}[!t]
\centering
\vspace{-0.35cm}	
\subfigtopskip=2pt
\subfigbottomskip=2pt
\subfigcapskip=-5pt
\setlength{\abovecaptionskip}{2pt}

\subfigure[Proposed candidates: Covariates in $\Delta_0$ are not used in model $m$, $m> 1$. ]{
\includegraphics[width=0.9\linewidth]{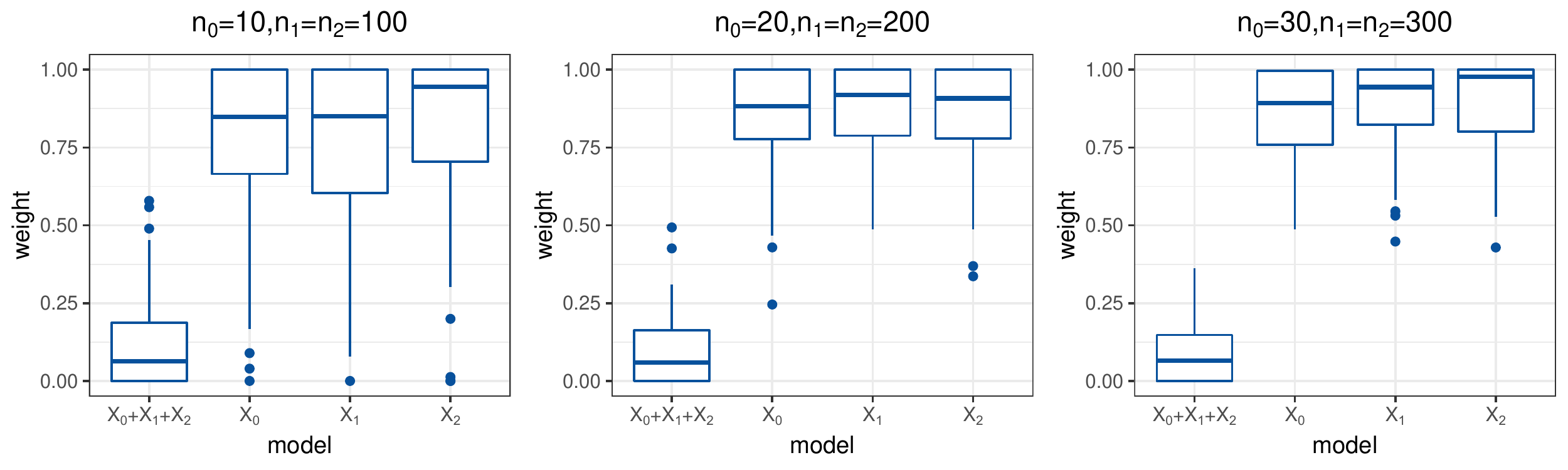}}	

\subfigure[Alternative candidates: Covariates in $\Delta_0$ are repeatedly used in model $m$, $m\geq 1$. ]{
\includegraphics[width=0.9\linewidth]{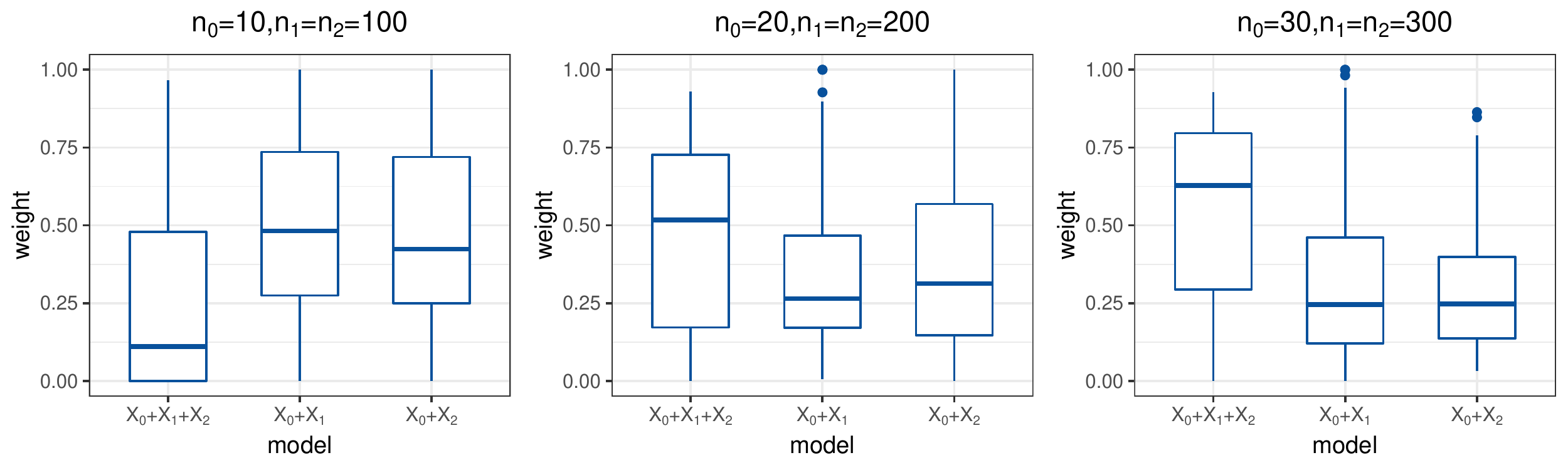}}	

\caption{An illustrative example. Weights assignment under finite sample cases.}
\label{fig:example}
\end{figure}

In existing literatures, incomplete cases data have also been utilized to construct candidate models but with covariates in $\Delta_0\cup\Delta_m$ \citep{Xiang2014, Fang:2017}. They differ significantly from our candidate estimators by using the covariates in $\Delta_0$ repeatedly, while $\hat{\mu}(\bx; \bD_{m}),1\leq m \leq M+1$ are estimated based on $M+1$ incomplete cases data-blocks with distinct covariates. Here, we use an illustrative example to provide some insights into the proposed approach.

\noindent{\bf An illustrative example}.  Assume that the true regression function is linear with independent covariates:
\[
\mu(\bX)=\beta_0 X_0+\beta_1 X_1+\beta_2 X_{2},
\]
where $\beta_0=\beta_1=\beta_{2}=1$ and the error term $\varepsilon \sim N(0,1)$.  Suppose that $\Delta_j=\{j\}$, $j=0,1,2$. We consider four linear candidate models with least squares estimation in the proposed method. It is true that $\hat{\mu}(\bx;\bD_0)$ is unbiased but with larger variance due to smaller $n_0$, while $\hat{\mu}(\bx;\bD_m)$, $m\geq 1$ are biased with smaller variance due to larger $n_m$. Intuitively, based on these candidates, the signal in the regression function can be completely recovered by setting $w_0\rightarrow 0, w_m\rightarrow 1$ for $m\geq 1$. But when $\Delta_0$ is involved in all candidate models, the biases of estimators can be canceled out only when $w_0\rightarrow 1, w_m\rightarrow 0$ for $m\geq 1$, hence inducing higher variance due to the smaller sample size of the complete cases. In Figure \ref{fig:example}, we present the result of selected weights based on the two different candidates sets. When $n_0=20,n_1=n_2=200$, the risk of the combined estimators based on the proposed candidates and the alternative candidates are 0.088(0.120) and 0.268(0.203) respectively. This simulation result clearly highlight necessity of separating $\Delta_0$ from the other blocks in the construction of the candidate models and the benefit of allowing the weights to be unconstrained in sum.
\subsection{Weight Choice Criterion}

In the literature of model averaging, the weight vector is usually restricted such that $\sum_{m=0}^{M+1}w_m=1$. As explained earlier, in our context, the weight constraint may severely hinder the potential of the proposed model averaging estimator. Let $\bw=(w_0,w_1,\ldots, w_{M+1})^{T}$ belong to an enlarged weight set $Q$, where
\[
Q=\left\{\bw \in[0,1]^{M+2} : 0 \leq w_{m} \leq 1\right\}.
\]
Such relaxed weight constraints have been studied in the literature in different context \citep{Wang2014, Ando2017a}.
Then, the main concern is how to select appropriate weights from $Q$ so that the aggregated estimator  $\hat{\mu}(\bw)=\sum_{m=0}^{M+1}w_m\hat{\mu}(\bx; \bD_m)$ has a superior performance.

For revealing the weight choice criterion, we introduce some notations. Define
$\bH_0=\bX_{S_0,\cup_{m=0}^M\Delta_m}(\bX_{S_0,\cup_{m=0}^M\Delta_m}^{T}\bX_{S_0,\cup_{m=0}^M\Delta_m})^{-1}\bX_{S_0,\cup_{m=0}^M\Delta_m}^{T}$, and $\tilde{\bH}_0=\bG(\bH_0-\bI_{n_0})+\bI_{n_0}$, where $\bG=\diag\left((1-h_{1})^{-1},\cdots,(1-h_{n_0})^{-1}\right)$, and $h_{k}$ is the $k$-th diagonal element of $\bH_0$. Suppose $\bH_0$ is full rank. Let $\tilde{\bH}_1$ denote an $n_0 \times (\sum_{m=1}^{M}n_m)$ matrix with the $i$-th row $\bpsi_m^{T}(\bx_{S_{0}^{i}\Delta_0};\bX_{\bigcup_{m=1}^MS_m,\Delta_0})$. Similarly,
 $\tilde{\bH}_{m+1}$, $m \geq 1$ denotes an $n_0 \times n_m$ matrix, the $i$-th row of which is $\bpsi_{m+1}^{T}(\bx_{S_{0}^{i}\Delta_m};\bX_{S_m\Delta_m})$, where $\bx_{S_{0}^{i}\Delta_m}^T$ is the $i$-th row of $\bX_{S_{0}\Delta_m}$. Now for the complete cases, the prediction for $\mu(\bX)$ using aggregated estimation over $M+2$ candidates is given by
 \[
 \tilde{\bmu}_{S_0}(\bw)=w_0\tilde{\bH}_0\by_{S_0}+w_1\tilde{\bH}_1\by_{\bigcup_{m=1}^MS_m}
 +\sum_{m=2}^{M+1}w_m\tilde{\bH}_m\by_{S_{m-1}},
 \]
 which is a linear combination of predictive values based on $M+2$ candidates. It is important to note that the delete-one cross validation technique is applied for the candidate estimator $\hat{\mu}(\bx; \bD_0)$ since it involves all covariates with complete cases. Then, we can select the optimal weight by minimizing the prediction error on the complete cases, i.e,
\begin{equation}\label{eq:criterion}
  \hat{\bw}=\arg \min _{\bw\in Q}\left\|\by_{S_0}-\tilde{\bmu}_{S_0}(\bw)\right\|^2,
\end{equation}
where $\|\cdot\|$ denotes the Euclidean norm.
The criterion can be viewed as a measurement of prediction accuracy on the complete cases for the aggregated estimator, where the relative contribution of each candidate estimator is assessed by its predictability to the response variable $\by_{S_0}$. Furthermore, the criterion can make full use of all available data when constructing candidate models and selecting weight, thus to be expected to substantially improve the prediction accuracy. With the selected weight $\hat{\bw}$, the final aggregated estimation for regression function is
\begin{equation}
  \hat{\mu}(\hat{\bw})=\sum_{m=0}^{M+1}\hat{w}_m\hat{\mu}(\bx; \bD_m).
\end{equation}

\section{Theoretical results}\label{sec:property}
In this section, we first study the optimality of the selected weight and then consider several scenarios of regression function to compare the optimal risks of the proposed method with alternatives.

\subsection{Asymptotic optimality}\label{subsec:asymptotic}

To evaluate the optimality of the proposed method, we define the squared error loss function
\begin{equation*}\label{eq:loss}
  L_n(\bw)=\left\|\bmu_{S_0}-\hat{\bmu}_{S_0}(\bw)\right\|^2,
\end{equation*}
and the corresponding risk
\begin{equation*}
  R_n(\bw)=E\left\{L_n(\bw)\right\},
\end{equation*}
 where $\bmu_{S_0}$ is an $n_0$-dimensional vector with $i$-th element $\mu(\bx_{S_{0}^i,\cup_{m=0}^M\Delta_m})$, where $\bx_{S_{0}^i,\cup_{m=0}^M\Delta_m}^{T}$ is the $i$-th row of $\bX_{S_{0},\cup_{m=0}^M\Delta_m}$. And $\hat{\bmu}_{S_0}(\bw)$ is defined as the fitted value of $\bmu_{S_0}$ based on (\ref{eq:estimatortrain}), which is an $n_0$-dimensional vector with $i$-th element $\sum_{m=0}^{M+1}w_m\hat{\mu}(\bx_{S_{0}^i,\cup_{m=0}^M\Delta_m}^{T}; \bD_m)$.


\begin{theorem}\label{theo:asyopt}
  Assume the data is generated from model (\ref{eq:model}).  Let $\zeta_n=\inf_{\bw\in Q_n}R_n(\bw)$ and $\bar{n}_M=\sup_{1\leq m\leq M}n_m$. Suppose
  \begin{description}
    \item[(C1)] $\sup_{i\geq 1}E(\varepsilon_i^4) < \infty$,
    \item[(C2)] $\frac{1}{p}\bar{\lambda}\{\bH_0\}=O(n_0^{-1})$,
    \item[(C3)] $\sup_{m\geq 1}\lambda_{\max}(\tilde{\bH}_m\tilde{\bH}_m^{T})=O(1)$,
    \item[(C4)] $\frac{p^2}{n_0\zeta_n^2}\to 0$ and $\frac{M^6p\bar{n}_M}{\zeta_n^2}\to 0$,
  \end{description}
  where $\bar{\lambda}(\cdot)$ denotes the maximum diagonal element of a matrix and $\lambda_{\max}(\cdot)$ denotes the maximum eigenvalue. Then, we have
  \begin{equation}\label{eq:asymeff1}
    \frac{L_n\left(\hat{\bw}\right)}{\inf_{\bw\in Q}L_n\left(\bw\right)}\to_p 1,
  \end{equation}
  where $\rightarrow_p$ denotes convergence in probability as $n \rightarrow \infty$.
\end{theorem}

This theorem provides a theoretical foundation for the proposed procedure to optimally
combine each candidate model under the  relaxed weight constraint. It shows that, under
some mild conditions, the smallest $L_2$ loss (which is, of course, infeasible to achieve
because the true $\mu$ is unknown), can indeed be reached by our procedure.

In this theorem,
Condition \textbf{(C1)} concerns with the moment bound of the error term and
is satisfied by Gaussian error. Condition \textbf{(C2)} excludes extremely unbalanced design matrices in complete cases data. With full observed data, a similar condition has been assumed in \cite{Li1987}. Condition \textbf{(C3)} concerns with SQD data, which requires that $\tilde{\bH}_m$ can behave as a projection matrix if $n\to \infty$, although $\tilde{\bH}_m, m\geq 1$
are not projection matrixes indeed. Specially, if all candidate estimators are least squares and $n_0=O(n_m)$, it can be easily satisfied when $n_0^{-1}\bX_{S_0\Delta_m}^T\bX_{S_0\Delta_m}$ and
$n_m^{-1}\bX_{S_m\Delta_m}^T\bX_{S_m\Delta_m}$ converge to a same matrix $\bOmega_m$.
A prerequisite for Condition \textbf{(C4)} to hold is $\zeta_n \to \infty$, which has been
commonly assumed in the literature \citep{Li1987,hansen07,HansenandRacine2012}.
As \cite{hansen07} remarked, this prerequisite specifies that there is no
candidate model for which the approximation bias is zero, this is conventional
for nonparametric regression. More specifically, when $p$ and $M$ are fixed, Condition \textbf{(C4)} just sets an upper bound on the rate of $\bar{n}_M$ by $\zeta_n$, which can be satisfied when $\bar{n}_M$ is a constant multiple of $n_0$. More detailed discussion about this condition can be found in \cite{Wanetal10}, \cite{Ando2014} and \cite{Fang:2017}.

\begin{corollary}\label{corollary1}
  Under the same assumptions in Theorem~\ref{theo:asyopt}, we have
  \begin{equation}\label{eq:asymeff2}
    \frac{R_n(\hat{\bw})}{\inf_{\bw\in Q}R_n\left(\bw\right)}\to_p 1.
  \end{equation}
  If in addition, $\left[L_n(\bw)-\zeta_n\right]\zeta_n^{-1}$ is uniformly integrable, then
  \begin{equation}\label{eq:asymeff3}
    \frac{EL_n(\hat{\bw})}{\inf_{\bw\in Q}R_n\left(\bw\right)}\to 1.
  \end{equation}
\end{corollary}

Corollary~\ref{corollary1} gives two similar statements about asymptotic optimality, similar results have been derived in \cite{HansenandRacine2012} and \cite{Zhang2020Parsimonious}. This corollary shows that the proposed procedure is asymptotically optimal in the sense that its risk is asymptotically identical to that of the infeasible but best possible averaged model.

The asymptotic optimality with the flexible weights show that our estimator can perform well adaptively in different scenarios. It is conceivable that sometimes the CC method may actually be the best and SQUARE achieves its performance with $w_0=1, w_m=0$ for $m\geq 1$; when the regression function can be well approximated by e.g., a simple combination of the incomplete cases models and SQUARE achieves it with $w_0=0, w_m=1$ for $m\geq 1$.

\subsection{Comparison between SQUARE and CC method}\label{subsec:comparisons1}

The general asymptotic optimality established in Corollary~\ref{corollary1} allows an examination of the performance of SQUARE in comparison to other methods in specific situations. To that end, we just need to compare the optimal risk $\inf_{\bw\in Q}R_n\left(\bw\right)$ with the risks of other methods. In this subsection, we study the difference of risks in two possible scenarios when true regression function is linear and nonlinear respectively.

\subsubsection{Linear regression}\label{subsub:Linear}

Suppose the data is generated from (\ref{eq:model}) with $\mu(\bX)=\sum_{j=1}^{p}X_j\beta_j$. To meet the condition $\zeta_n \to \infty$, we assume that the common set $\Delta_0=\varnothing$ and the covariates in subsets $\Delta_1$ and $\Delta_2$ are used to estimate $\mu(\bX)$ while the remaining covariates in $\Delta_3=\{1,\ldots,p\}\setminus \Delta_1\cup\Delta_2$ are excluded in our estimation procedure, where $\Delta_1\cup\Delta_2 \subset \{1,\ldots,p\}$, $\Delta_1\cap\Delta_2=\varnothing$, $|\Delta_1|=p_1$, $|\Delta_2|=p_2$. The signal in each subset is denoted by $s_k=\sum_{j\in \Delta_k}\beta_j^2>0$, $k=1,2,3$. We assume that the sample size of each data-block satisfies $n_1=n_2, n_0=o(n_1)$. Suppose that all candidate models are estimated by least squares. Define the projection matrixes
$\bH_0=\bX_{S_0,\Delta_1\cup\Delta_2}\left(\bX_{S_0,\Delta_1\cup\Delta_2}^T\bX_{S_0,\Delta_1\cup\Delta_2}\right)^{-1}\bX_{S_0,\Delta_1\cup\Delta_2}^T,$~
$\bH_m=\bX_{S_0\Delta_m}\left(\bX_{S_m\Delta_m}^T\bX_{S_m\Delta_m}\right)^{-1}\bX_{S_m\Delta_m}^T,$
for $m=1,2.$
Then the risk function of SQUARE is
\begin{equation}\label{eq:risksquare}
\begin{split}
  R_1(\bw)=\left\|\bmu_{S_0}-\sum_{m=0}^{2}w_m\bH_m\bmu_{S_m}\right\|^2+w_0^2(p_1+p_2)\sigma^2+\sum_{m=1}^{2}\frac{n_0}{n_m}w_m^2p_m\sigma^2.
\end{split}
\end{equation}

To simplify the calculations, in our illustration, we consider further that $n_m^{-1}\bX_{S_m\Delta_k}^T\bX_{S_m\Delta_l}=\delta_{kl}\bI_{kl}$ for $m=0,1,2,$ and $k,l=1,2,3$, where $\delta_{kl}$ is the Kronecker delta and $\bI_{kl}$ is a $p_k\times p_l$ diagonal matrix with
1 in position $(i, i)$. Elementary algebraic calculations show that the optimal weights are $w_0^{opt}=0, w_1^{opt}=w_2^{opt}=1$, and thus the resulting optimal aggregated estimator has the same rate of convergence with that of the least square estimator with $n_1$ complete observations. Since $n_0=o(n_1)$, we see clearly that SQUARE performs much better in risk reduction than CC method in this setting.

Now, let us consider another case where two blocks share some common variables, i.e, $\Delta_1 \cap \Delta_2=\Delta^*$ and $s^*=\sum_{j\in \Delta^*}\beta_j^2>0$, which implies there exist a strong correlation between the two subsets due to $\Delta^*$. With some similar algebraic calculations, we have the optimal weights $w_0^{opt}=1, w_1^{opt}=w_2^{opt}=0$. Thus, in this case, CC method is the best among the weights considered. An important point is that in those cases, SQUARE automatically achieves the best performance.

\subsubsection{Nonlinear regression}

Suppose the true regression function is $\mu(\bX)=\mu_1(X_1)+\mu_2(X_2)$, where $\mu_1$ and $\mu_2$ are mutually orthogonal and belong to a Sobolev class
$$\mathcal{F}(\alpha,L)=\left\{\mu\in[0,1]\to \mathbf{R}:\mu^{(\alpha-1)}\;\text{is absolutely continuous and} \int_{0}^{1}(\mu^{(\alpha)}(x))^2dx\leq L^2  \right\}.$$
In this case, $\Delta_1=\{1\}$ and $\Delta_2=\{2\}$. Define two incomplete data-blocks $\bX_{S_m\Delta_m},m=1,2$ and a collection of basis functions $\{\phi_1, \ldots, \phi_k, \ldots \}$, examples are tensor product splines basis of different orders and varying number and locations of knots, or wavelet basis with different resolutions. Denote $I_m=\{i_{m,1}, i_{m,2},\ldots,i_{m,k_m}\}\in \Gamma$, where $\Gamma$ is a finite subset of $\{1,2,\ldots\}$, $\bPhi_{I_m}=\left(\phi_{i_{m,1}}(\bX_{S_m\Delta_m}),\ldots,  \phi_{i_{m,k_m}}(\bX_{S_m\Delta_m})\right)$ for $m=1,2$. We construct two candidate models $\hat{\mu}_{m,I_m}(x)=\sum_{j=1}^{k_m}\hat{\theta}_{i_{m,j}}\phi_{i_{m,j}}(x),m=1,2$, where $\hat{\theta}_{i_{m,j}}$ is the $j$-th element of least squares estimator $(\bPhi_{I_m}^T\bPhi_{I_m})^{-1}\bPhi_{I_m}^T\by_{S_m}$. We use the ABC criterion \citep{Yang:1999} to select $I_m,m=1,2$ in practice. In this situation, the risk function has the same form with (\ref{eq:risksquare}) except that $\bH_m=\bPhi_{I_m}(\bPhi_{I_m}^T\bPhi_{I_m})^{-1}\bPhi_{I_m}^T$ and $p_m=k_m$.

Under the orthogonality assumption of $\bH_m,m=1,2$, $\hat{\mu}_{m,I_m}(x)$ converges to $\mu_m(x)$ in the rate of $n_m^{-2\alpha/(2\alpha+1)}$ \citep[see][Theorem 5]{Yang:1999}, but cannot converge to $\mu(x)$ since it misses a part of the signal. It is also expected that the optimal weights are $w_0^{opt}=0, w_1^{opt}=w_2^{opt}=1$, since the complementary signals in $\hat{\mu}_{m,I_m}(x),m=1,2$ can be exactly combined and their biases can be canceled out by assigning them the weight of one. Thus optimal rate for risk of the resulting model averaging estimator has order $n_1^{-2\alpha/(2\alpha+1)}$, which is much smaller than the non-converging risk of least squares estimator based on CC data in the nonlinear setting.

\subsection{Relaxation of weight constraint}\label{subsec:comparisons2}

It is important for SQUARE to allow the weights to vary freely between 0 and 1 without the restriction of summing up to 1. We find that this relaxation does give us more potential in risk compared to model averaging estimators with the standard weights constraint, such as the method proposed by \cite{Fang:2017}, which uses delete-one cross-validation based on CC data to determine the optimal weight vector and is denoted as ``CC-JMA'' here. In particular, we have the following result under the linear regression setup given in subsection \ref{subsub:Linear}.

\begin{proposition}\label{lem:comparison}
  Let $R_{1}^{\ast}=\inf_{\bw \in Q_1}R_{1}(w_0, w_1, w_2)$ and $R_{2}^{\ast}=\inf_{\bw \in Q_2}R_{2}(w_0, w_1, w_2)$ denote the infeasible optimal risk of the proposed estimator and the model averaging estimator of \cite{Fang:2017} respectively, where $Q_1=\left\{(w_0,w_1,w_2)^T \in[0,1]^{3} : 0 \leq w_{m} \leq 1\right\}$ and $Q_2=\{(w_0,w_1,w_2)^T \in[0,1]^{3} : w_0+w_1+w_2=1\}$. \\
\textbf{Scenario 1}. If $s_1>0$ and $s_2=0$, we have $R_{1}^{\ast}-R_{2}^{\ast}=o(1)$. \\
\textbf{Scenario 2}. If $s_1>0$ and $s_2>0$, we have $R_{1}^{\ast} < R_{2}^{\ast}$ and $R_{2}^{\ast}-R_{1}^{\ast} \sim (p_1+p_2)\sigma^2$, where $a_n \sim b_n$ means $\lim a_n/b_n \to 1$. \\
  Moreover, in these two different scenarios, SQUARE converges at the same rate.
\end{proposition}

In Proposition~\ref{lem:comparison}, we theoretically justifies that the relaxation of weights can provide substantial advantages in risk when candidate models are estimated by different data-blocks of SQD data. Specifically, under scenario 1 where only one subset has nonzero signal, the difference between the minimal risks of SQUARE and CC-JMA can be neglected asymptotically. Under scenario 2 where both two subsets contain important variables, however, the improvement of the proposed method over CC-JMA is substantial. In addition, the risk of SQUARE converges at the same rate in both scenarios, which implies that the proposed estimator is more robust while CC-JMA performs poorly in the second scenario. More advantages of SQUARE can be found in our numerical studies.

\section{Simulation}\label{sec:simulation}

In this section, we conduct simulation studies to evaluate performance of SQUARE and compare it with alternatives for SQD-type data. We focus on the infinite-order regression setting and simulations involving nonlinear regression are provided in the Supplementary Material due to space limitation.

To better gauge performance of the proposed approach, we consider the following alternatives. The first is CC method that only uses samples with full observed covariates. The second is multiple imputation (MI) method for analyzing missing data, which can be implemented in the R package \verb"mice" with default settings \citep{mice}. The third is
 Mallows model averaging (IMP-MMA) proposed by \citep{Zhang:2013}, under which the missing values are replaced by zeros and the weights are determined by minimizing a Mallows $C_p$ criterion. The last one is CC-JMA \citep{Fang:2017}, which performs weight selection by delete-one cross-validation on CC data with the restriction of summing up to 1. Additionally, it differs from SQUARE by fitting candidate models using common sets repeatedly.

\subsection{Simulation Setup}
We consider a linear regression model
\begin{equation}\label{eq:datagnr2}
  Y=\mu(\bX)+\varepsilon=\sum_{j=1}^{1000}X_{j}\beta_j+\varepsilon,
\end{equation}
where $\varepsilon$ is the random error generated from $N(0,\sigma^2)$. Here, the variance $\sigma^2$ is set to make $R^2=\mbox{var}(\mu(\bX))/\mbox{var}(Y)$ vary between 0.1 and 0.9. Only the first $p$ covariates $\bX_{\Delta}=(X_1,\cdots,X_p)^T$ are used when applying the SQUARE and  alternatives, where $\Delta=\{1,\cdots,p\}$. For $j>p$, $X_j$ is generated independently from normal distribution with mean 0 and variance 1 and $\beta_j=1/j$. For the first $p$ covariates, we consider different modular structures with different module sizes. Structure I: $p=28,~ M=5,~ p_0=3,~ p_1=\ldots =p_M=5$; Structure II: $p=28,~ M=3,~ p_0=3,~ p_1=15,~ p_2=5,~ p_3=5$.


We generate the covariates $\bX_\Delta$ following the ``rule" that the covariates in the same subset are more correlated than the covariates in different subsets. Specifically, let $\bX_\Delta=(\bX_{0}^T,\bX_{1}^T,\ldots, \bX_{M}^T)^T$ and $\bbeta_\Delta=(\bbeta_0^T,\bbeta_1^T,\ldots, \bbeta_M^T)^T$, where $\bX_{m}$ is a $p_m$-dimensional vector which contains covariates in the subset $\Delta_m$ and $\bbeta_m$ is the corresponding coefficients vector. The intercept term is included in $\Delta_0$ by setting $X_1=1$. To generate remaining covariates, we first generate $\bZ$ from a $(p-1)$-dimensional normal distribution with mean $\mathbf{1}_{p-1}$ and covariance matrix $\bOmega$, where
\begin{equation*}
  \bOmega=\left(
  \begin{array}{cccc}
    \bOmega_{00}& \bOmega_{01}& \ldots& \bOmega_{0M}\\
    \bOmega_{10}& \bOmega_{11}& \ldots& \bOmega_{1M}\\
    \vdots & \vdots & & \vdots\\
    \bOmega_{M0}& \bOmega_{M2}& \ldots& \bOmega_{MM}\\
  \end{array}
  \right).
\end{equation*}
Here, $\bOmega_{00}$ and $\bOmega_{mm}$ ($m >1$) are $(p_0-1)\times (p_0-1)$ and $p_m \times p_m$ square matrices with diagonal elements equal to $1$ and off-diagonal elements equal to $\lambda_1$ and $\lambda_2$ respectively. For $m\geq1$, $k\geq1$ and $m \neq k$, $\bOmega_{m0}=\bOmega_{0m}^T$ is a $p_m \times (p_0-1)$ matrix with elements $\lambda_1$, $\bOmega_{mk}=\bOmega_{km}^T$ is a $p_m \times p_k$ matrix with elements $\lambda_3$. We set $\lambda_1=0.1, \lambda_2=0.3, \lambda_3=0.1$ to meet the ``rule". To generate $\bX_\Delta$, we set $X_i=1\{Z_i \geq 0.885\}$ for $i \in \left\{3,8,12,16,20,24\right\}$ and $X_i=Z_i$ otherwise, so that both discrete and continuous covariates are incorporated in our analysis.

We further consider several cases for coefficients $\bbeta_\Delta$:
\begin{itemize}
  \item[] Case 1: $(\bbeta_0^T,\bbeta_1^T,\ldots, \bbeta_M^T)^T=c_1\left(1,\frac{1}{3}\mathbf{1}_{p_0-1}^T,\frac{1}{3}\mathbf{1}_{p_1}^T,\frac{1}{3}\mathbf{1}_{p_2}^T,\ldots,\frac{1}{3}\mathbf{1}_{p_M}^T\right)^T$,
  \item[] Case 2: $(\bbeta_0^T,\bbeta_1^T,\ldots, \bbeta_M^T)^T=c_2\left(1,\frac{1}{3}\mathbf{1}_{p_0-1}^T,\mathbf{f}_{p_1}^T,\mathbf{f}_{p_2}^T,\ldots,\mathbf{f}_{p_M}^T\right)^T$,
  \item[] Case 3: $(\bbeta_0^T,\bbeta_1^T,\ldots, \bbeta_M^T)^T=c_3\left(1,\frac{1}{3}\mathbf{1}_{p_0-1}^T,\mathbf{g}_{p_1}^T,\frac{1}{2}\mathbf{g}_{p_2}^T,\ldots,\frac{1}{M}\mathbf{g}_{p_M}^T\right)^T$,
\end{itemize}
where $\mathbf{1}_{d}$ is an $d$-dimensional vector with each element equal to $1$, $\mathbf{f}_{d}$ and $\mathbf{g}_{d}$ are two $d$-dimensional vectors with the $k$-th element being  $\frac{1}{2(k-1)+1}$ and $\frac{1}{k}$ respectively, and $c_1$, $c_2$, $c_3$ are chosen such that $\mbox{var}(\mu)=10$ to facilitate comparisons between different cases.
 Case 1 just implies that all covariates are equally contributive. Case 2 mimics the situation that all modules are contributive, but the covariates in modules are not equally important. Case 3 means that both modules and the covariates in each module are not in the same importance.

The simulated SQD-type data is composed of $n_0$ randomly generated observations with complete information, and $n_m$ ($m\geq 1$) randomly generated cases with covariates in subset $\Delta_0$ and subset $\Delta_m$ being observed. For simplicity, we set $n_1=\cdots=n_{M}$, $n_0$ varies as 40, 50 and 60, and $n_1$ varies as 100, 150 and 200 respectively. For all settings, $B=1000$ replicates are simulated. When evaluating the proposed and alternative approaches, of the most interest is prediction accuracy. In each replication, we simulate an independent testing dataset with sample size $n_{test}=10000$ under the same settings but without missing. Then we compute the predicted values of $\boldsymbol{\mu}$ on the testing data, and the predication accuracy is measured by
\[
\mbox{MSE}=\frac{1}{n_{test}}\|\hat{\boldsymbol{\mu}}-\boldsymbol{\mu}\|^2,
\]
where $\hat{\boldsymbol{\mu}}$ is the estimate of $\boldsymbol{\mu}$ over the testing data. Furthermore, the mean of MSE over $B$ replicates can be decomposed into $variance$ and $bias^2,$ where
\begin{eqnarray*}
variance=\frac{1}{Bn_{test}}\sum_{b=1}^{B}\left\|\hat{\boldsymbol{\mu}}^{(b)}-\bar{\hat{\bmu}}\right\|^2,~~~
\mbox{and}~~~ bias^2 =\frac{1}{n_{test}}\left\|\bar{\hat{\bmu}}-\boldsymbol{\mu}\right\|^2.
\end{eqnarray*}
here $\bar{\hat{\bmu}}=\frac{1}{B}\sum_{b=1}^B\hat{\bmu}^{(b)}$ and $\hat{\bmu}^{(b)}$ is the estimate of $\bmu$ in the $b$-th replicate.
It means that the MSE has two components: one measures the variability of the estimator and the other measures the bias. An estimator that has good MSE property has small combined variance and bias.

\subsection{Simulation Results}
\begin{figure}[!t]
\centering
\vspace{-0.35cm}	
\subfigtopskip=2pt
\subfigbottomskip=2pt
\subfigcapskip=-5pt
\setlength{\abovecaptionskip}{2pt}

\subfigure[Case 1]{
\includegraphics[width=1\linewidth]{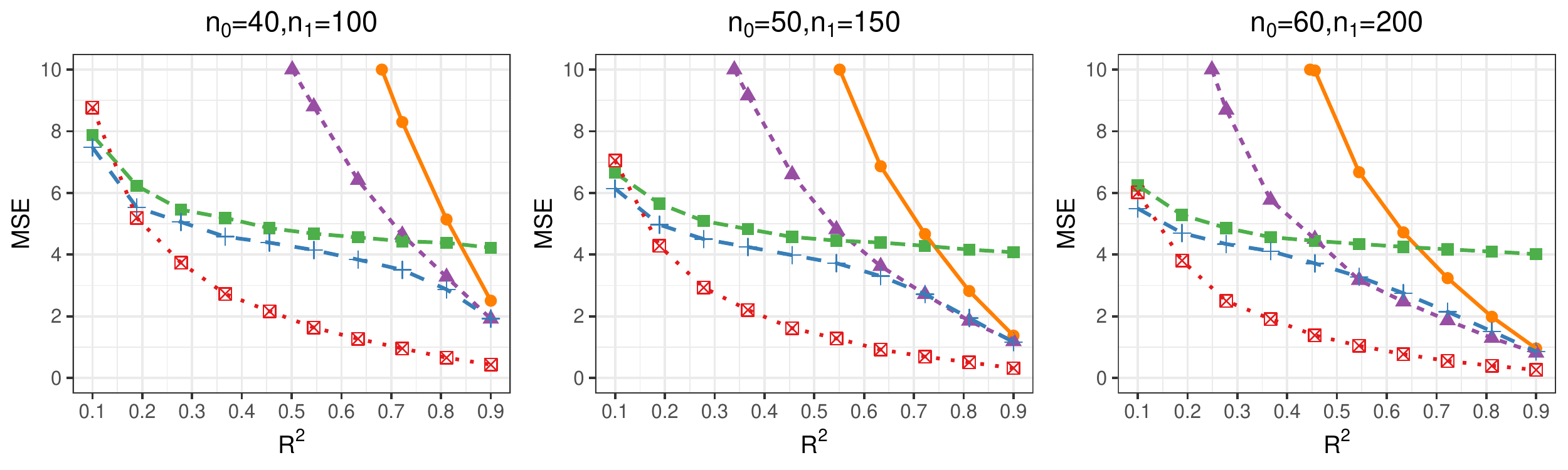}}	

\subfigure[Case 2]{
\includegraphics[width=1\linewidth]{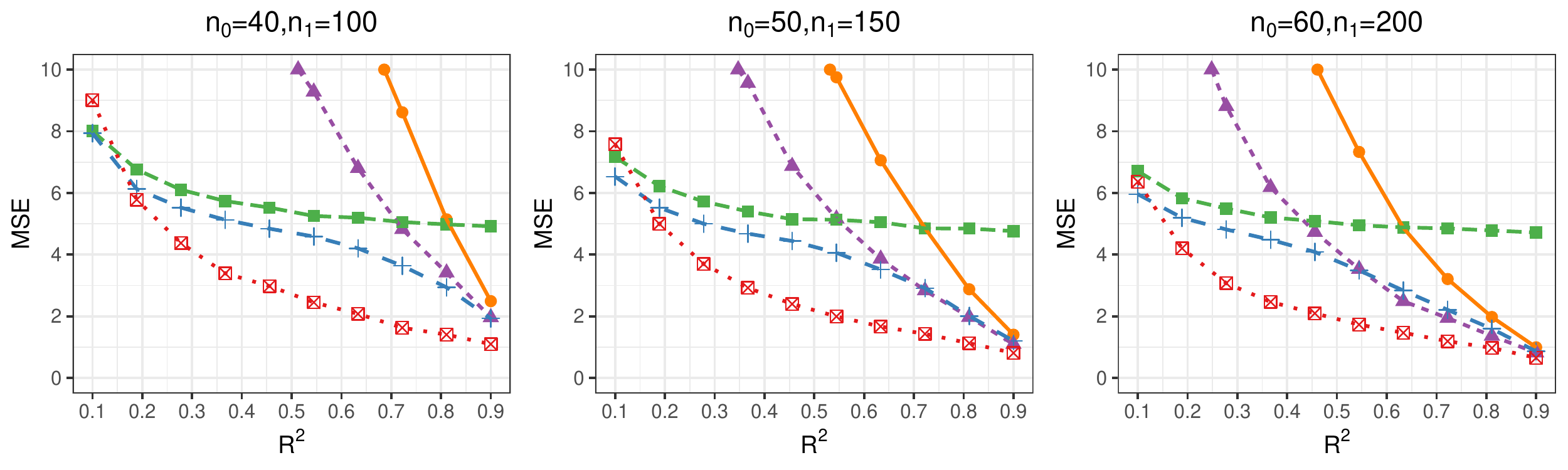}}	

\subfigure[Case 3]{
\includegraphics[width=1\linewidth]{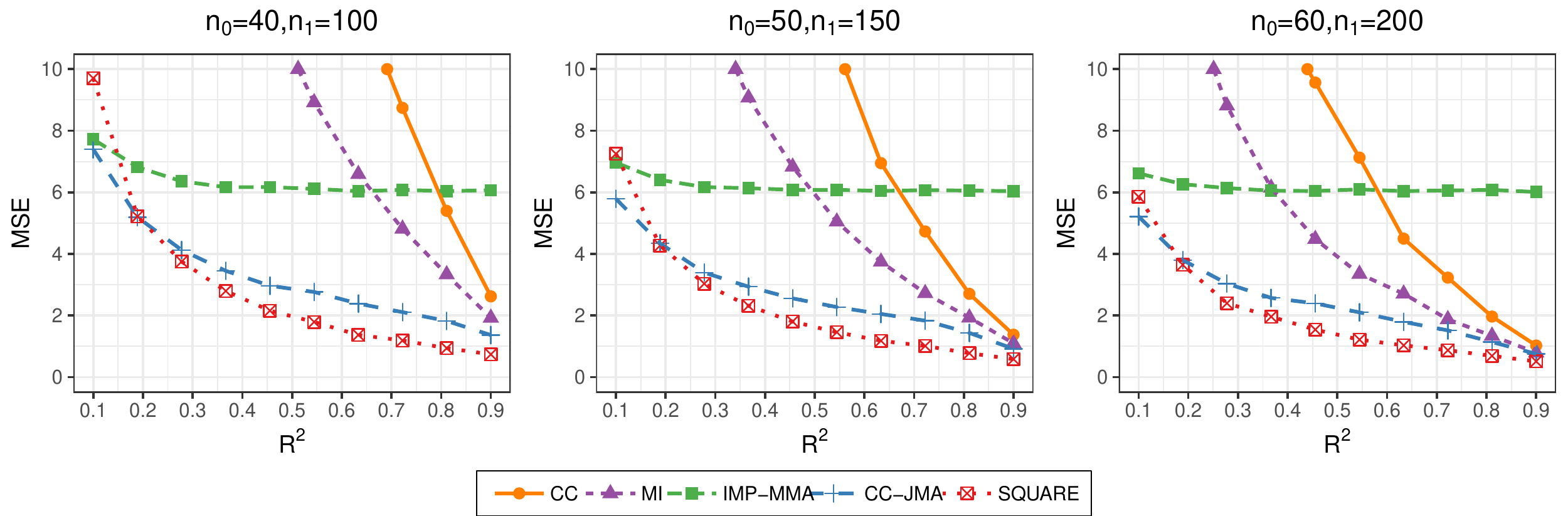}}	

\caption{Comparison of MSE among different methods under modular structure I.}
\label{fig:mse:s1}
\end{figure}

\begin{figure}[!t]
\centering
\vspace{-0.35cm}	
\subfigtopskip=2pt
\subfigbottomskip=2pt
\subfigcapskip=-5pt
\setlength{\abovecaptionskip}{2pt}

\subfigure[Case 1]{
\includegraphics[width=1\linewidth]{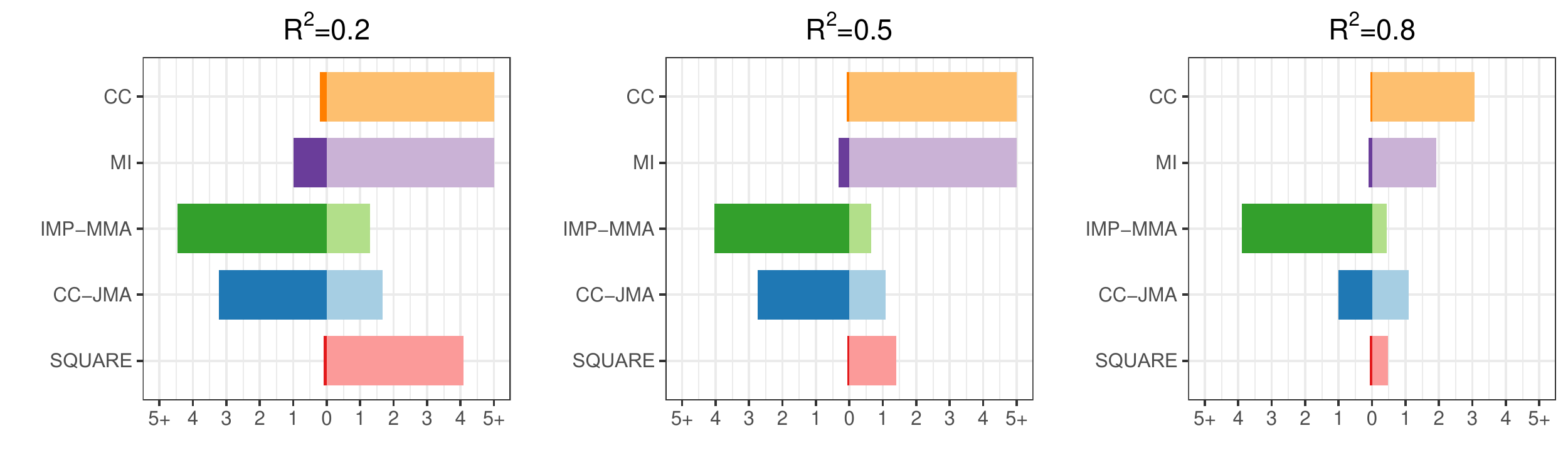}}

\subfigure[Case 2]{
\includegraphics[width=1\linewidth]{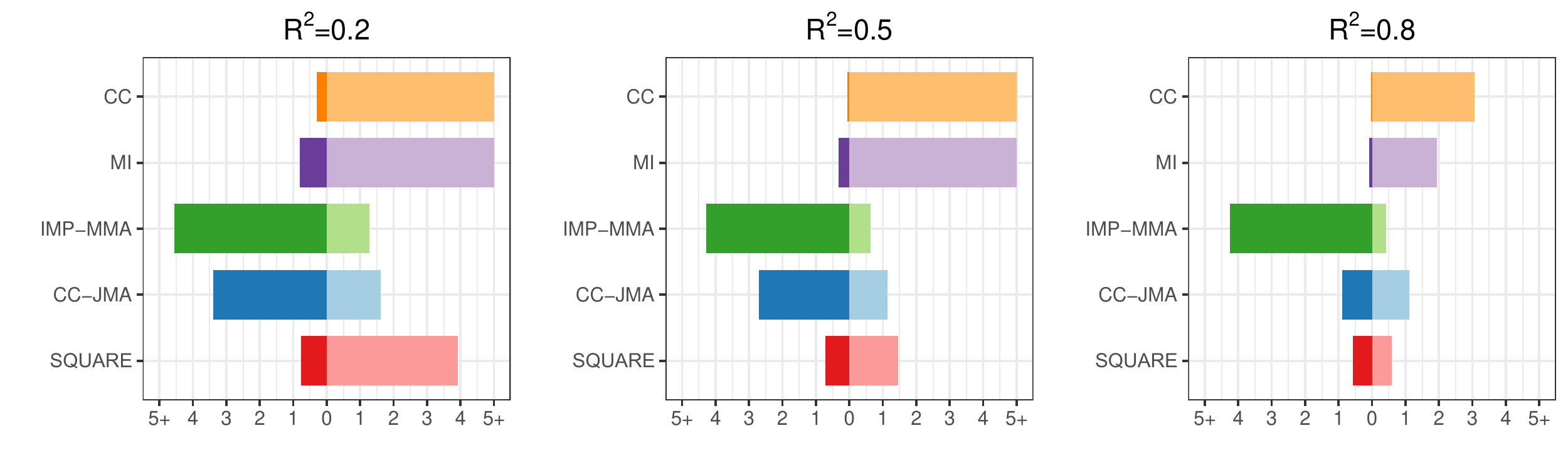}}

\subfigure[Case 3]{
\includegraphics[width=1\linewidth]{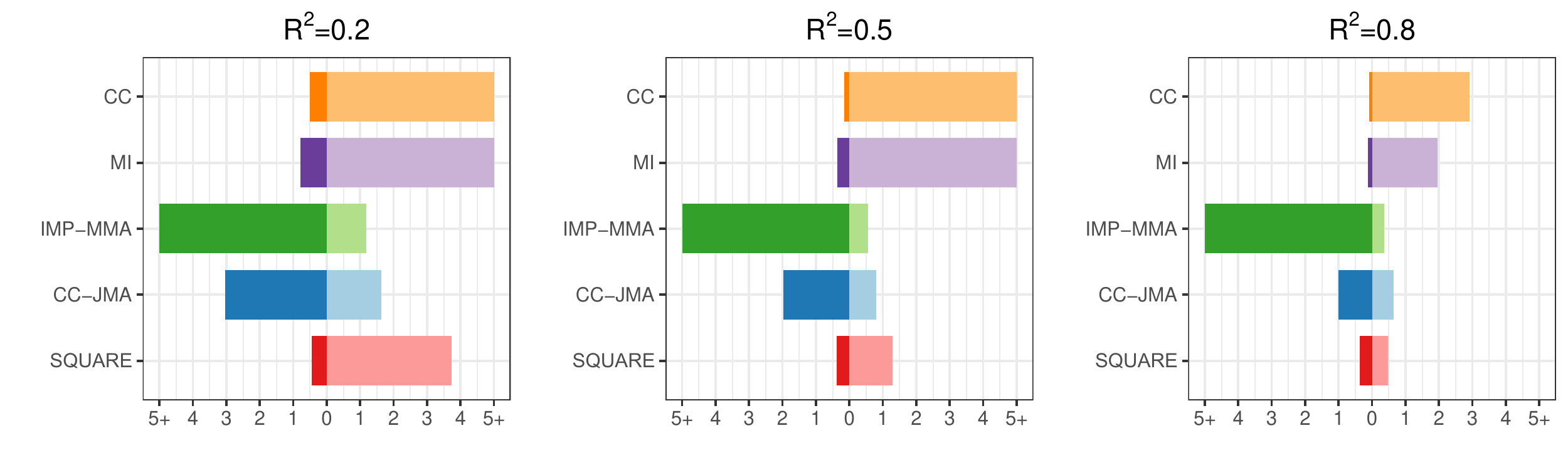}}

\caption{Comparison of bias (dark color) and variance (light color) among different methods under modular structure I with $n_0=50, n_1=150$.}
\label{fig:bv:s1}
\end{figure}

\begin{figure}[!t]
\centering
\vspace{-0.35cm}	
\subfigtopskip=2pt
\subfigbottomskip=2pt
\subfigcapskip=-5pt
\setlength{\abovecaptionskip}{2pt}

\subfigure[Case 1]{
\includegraphics[width=1\linewidth]{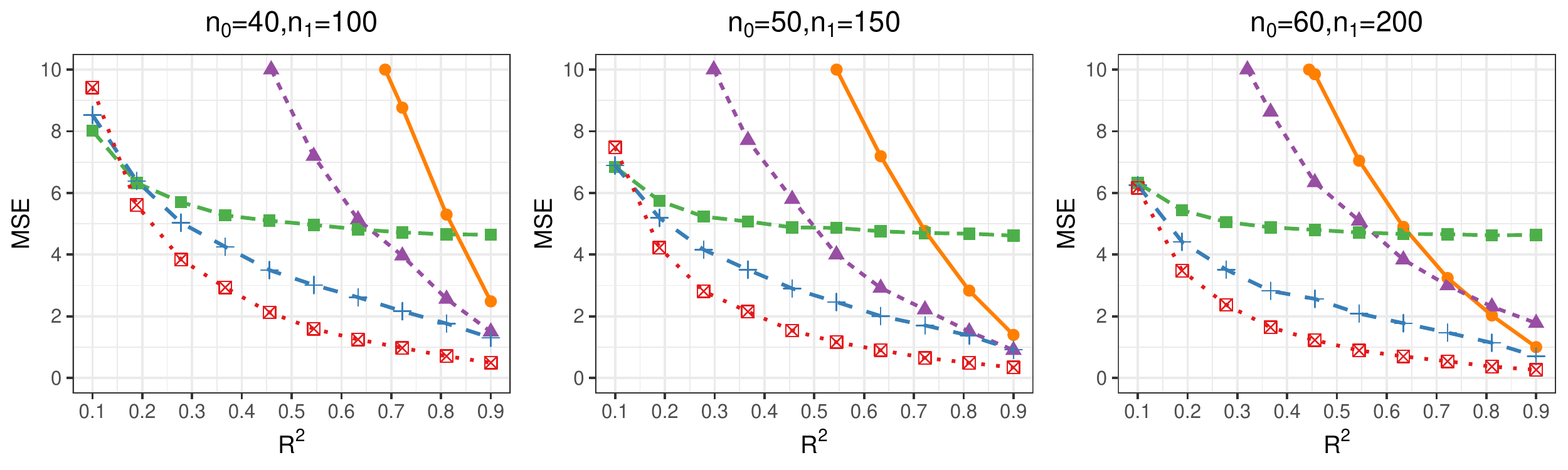}}	

\subfigure[Case 2]{
\includegraphics[width=1\linewidth]{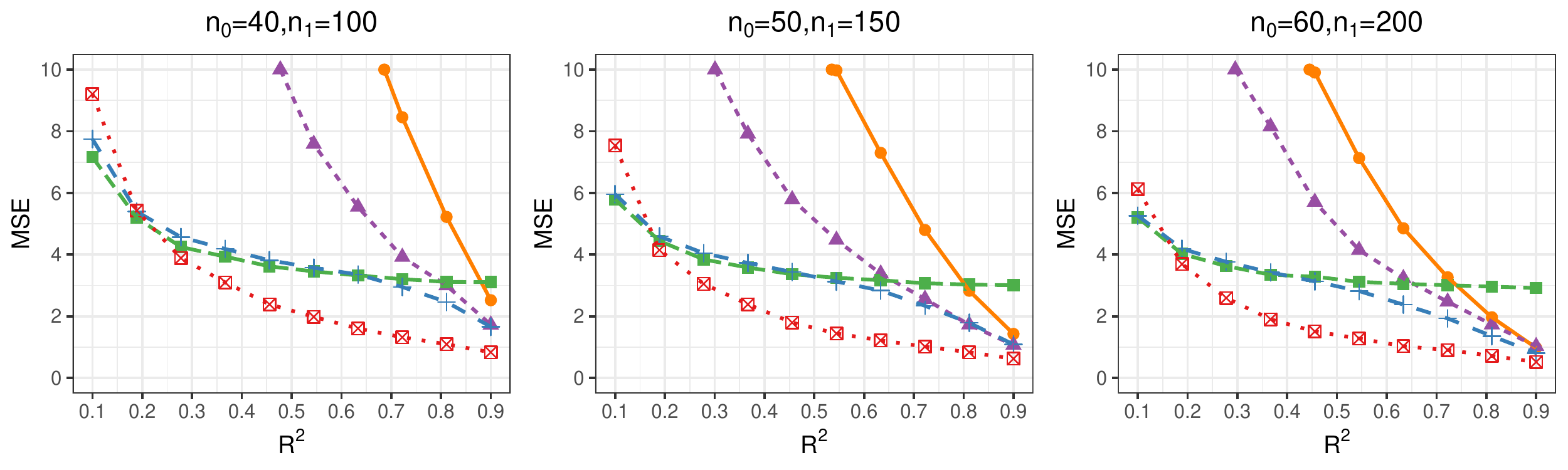}}	

\subfigure[Case 3]{
\includegraphics[width=1\linewidth]{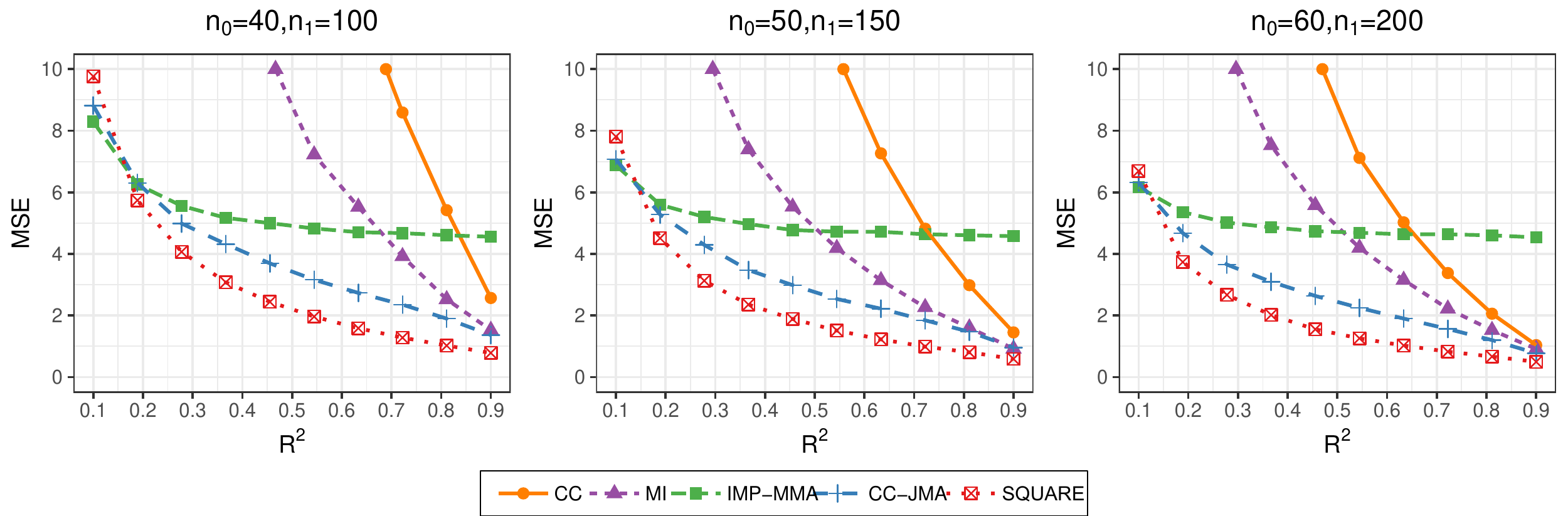}}	

\caption{Comparison of MSE among different methods under modular structure II.}
\label{fig:mse:s2}
\end{figure}

\begin{figure}[!t]
\centering
\vspace{-0.35cm}	
\subfigtopskip=2pt
\subfigbottomskip=2pt
\subfigcapskip=-5pt
\setlength{\abovecaptionskip}{2pt}

\subfigure[Case 1]{
\includegraphics[width=1\linewidth]{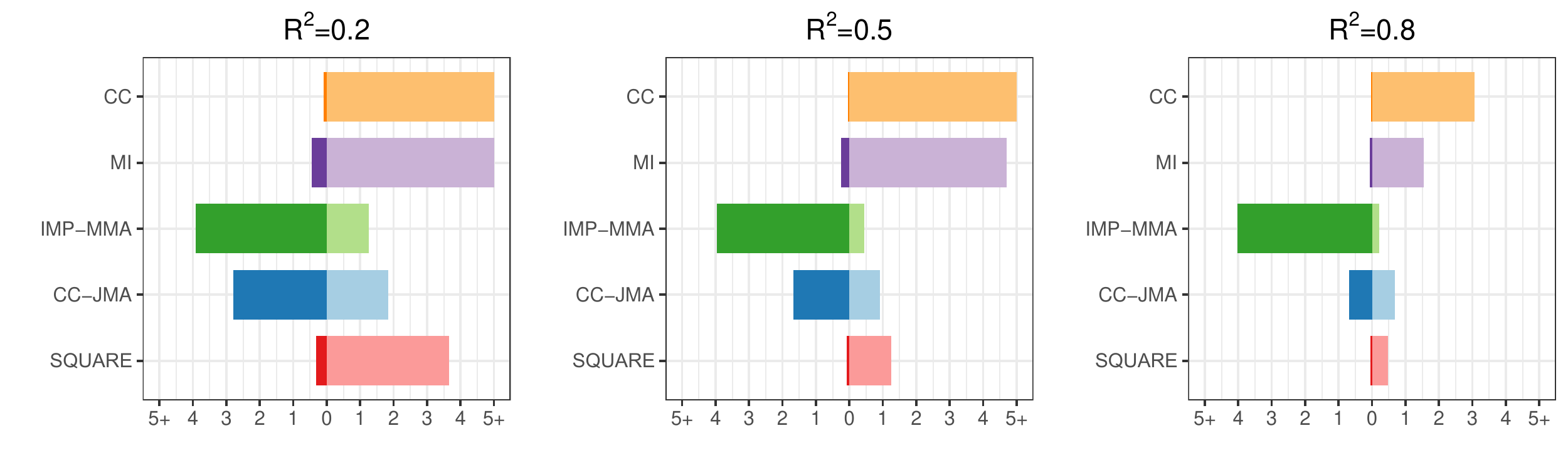}}	

\subfigure[Case 2]{
\includegraphics[width=1\linewidth]{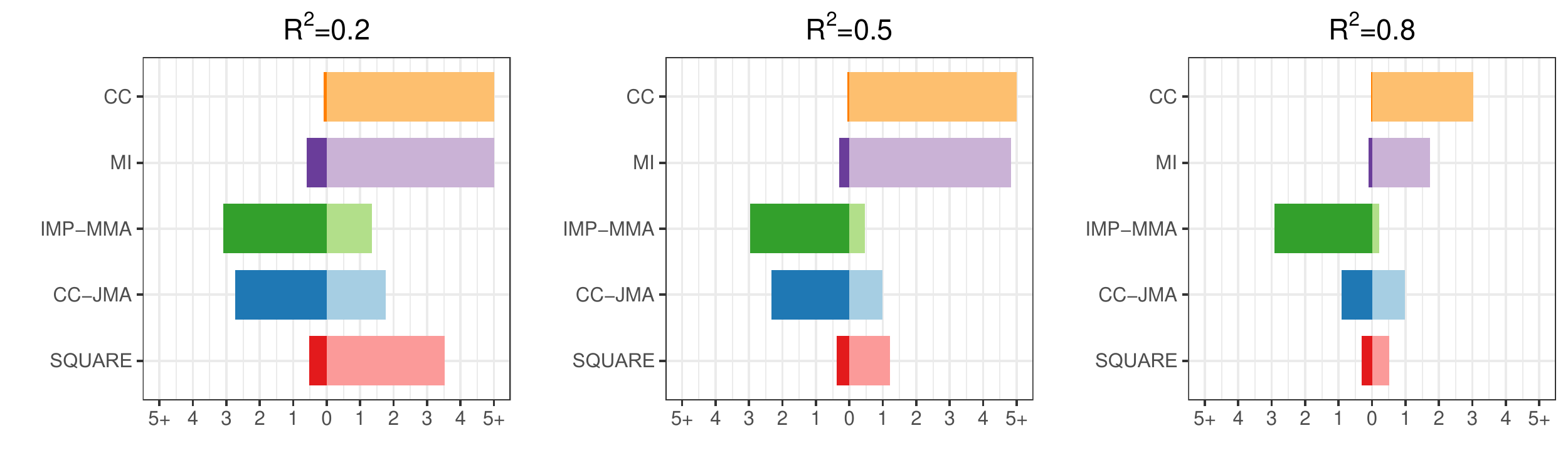}}	

\subfigure[Case 3]{
\includegraphics[width=1\linewidth]{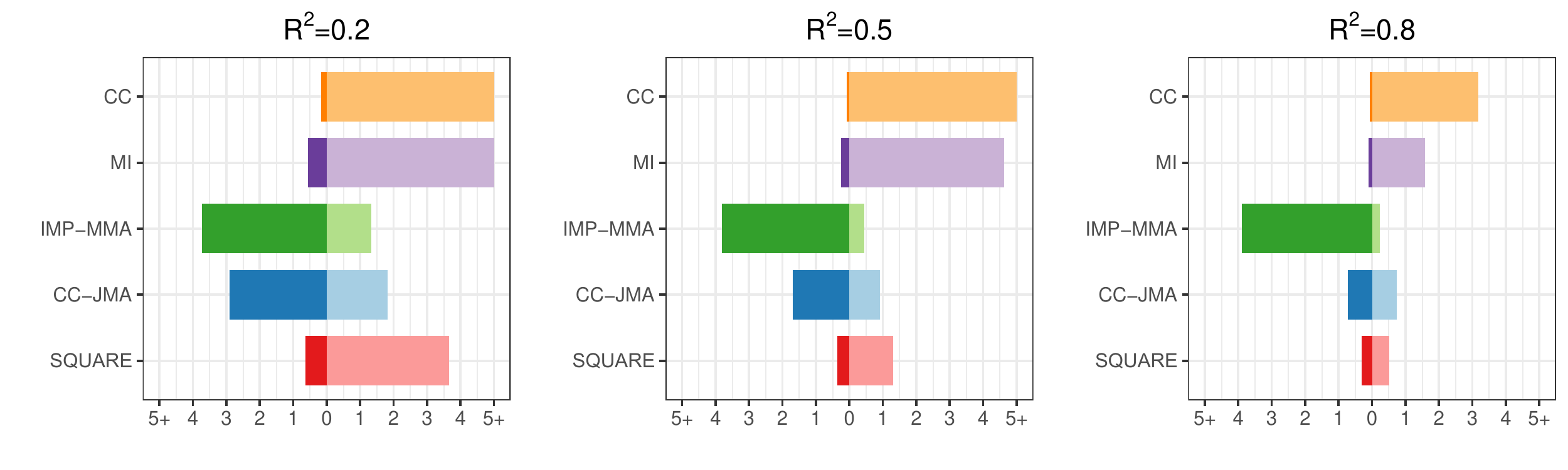}}	

\caption{Comparison of bias (dark color) and variance (light color) among different methods under modular structure II with $n_0=50, n_1=150$.}
\label{fig:bv:s2}
\end{figure}

The median of MSEs for modular structures I and II are displayed in Figures~\ref{fig:mse:s1} and \ref{fig:mse:s2} respectively. For $n_0=50, n_1=150$, we also provide a comparison of bias component and variance component of MSE in Figures~\ref{fig:bv:s1} and \ref{fig:bv:s2}. In each figure, the simulation results with three different cases of coefficients are displayed in rows (a), (b) and (c), respectively. The lines of CC and MI do not appear completely in some panels of figures, because these two methods lead to much larger MSEs than others when $R^2$ is small. The main conclusions are as follows.

From Figure~\ref{fig:mse:s1}, it is observed that the proposed method has a sizable edge over other methods basically across the range of $R^2$ under Cases 1 and 2. For example, when $n_0=50, n_1=150, R^2=0.5$, the medians of MSEs of CC, MI, IMP-MMA, CC-JMA, SQUARE are 11.05, 5.52, 4.51, 3.93, 1.30 and 10.23, 5.18, 5.13, 4.15, 2.05 under Cases 1 and 2 respectively. For Case 3, SQUARE doesn't have that great superiority over CC-JMA, which is consistent with the results in Proposition~\ref{lem:comparison}. Actually, the signals of Case 3 decreases in different subsets and this is closer to Scenario 1 of Proposition~\ref{lem:comparison}, where only one subset has nonzero signal.  However, Cases 1 and 2 are closer to Scenario 2 where all modules matters as depicted in the proposition. Nonetheless, SQUARE is somewhat more robust than the other two model averaging estimator as the cases of coefficients change. In addition, when $R^2$ is relatively small, SQUARE often approaches to or performs a little worse than IMP-MMA and CC-JMA in MSE. This is because that weight relaxation brings benefits in prediction accuracy, but also needs to pay more cost in selecting the optimal weight, especially when $R^2$ is small. As shown in Figure \ref{fig:bv:s1}, when $R^2=0.2$, the variance component of SQUARE dominates its MSE, which is much larger than that of IMP-MMA and CC-JMA. In most applications, however, we are not so hung-up about the case with small $R^2$.

Figure~\ref{fig:bv:s1} provides additional information on the composition of the bias and variance components of MSEs. Therefore, it as well provides a more comprehensive and in-depth comparison of competing methods. In general, we can observe an obvious trade-off between bias and variance of each method: the method with smaller bias tends to have higher variance than others, and vice versa. As shown in Figure~\ref{fig:bv:s1}, although CC and MI have lower bias, their variance becomes quite wild especially when $R^2=0.2$ and 0.5. For the three model averaging approaches,
we can always observe that IMP-MMA has the smallest variance but largest bias, resulting in less competitive performance in MSE. The reason is that IMP-MMA uses all $n_0+Mn_1$ samples to select weight, hence leading to a more stable estimate, but after first replacing the missing values with zeros, which may cause substantial bias of estimates in our settings.

The simulation results with modular structure II are presented in Figures~\ref{fig:mse:s2} and \ref{fig:bv:s2}. Different from structure I, the sizes of modules in structure II are different, as resembling that of the ESS data analyzed in Section~\ref{sec:realdata}. The results in Figure \ref{fig:mse:s2} show that SQUARE performs better than or comparable to the alternatives and similar superiority of the proposed approach is observed from Figure~\ref{fig:bv:s2} (details omitted).

In the supplementary material, we present two additional simulations: the first involves the linear regression function (\ref{eq:datagnr2}), but is designed to compare different methods under missing at random (MAR) setting. The second simulation involves nonlinear regression function and all candidate models are estimated based on B-spline function. Overall, these results also present evidence favoring the use of SQUARE with SQD-type data. Particularly, the advantage of the proposed method over CC-JMA is much more substantial under the nonlinear setting. Because CC-JMA only utilizes the complete cases data to build candidate models when calculating weights, this weights learning procedure may suffer quite high variability since more parameters need to be estimated under the nonlinear setting.

\section{ESS Data Analysis}\label{sec:realdata}

The European Social Survey (ESS) is an academically driven, large-scale social survey that was formally established in 2001. It has produced a large amount of rigorous data about people's underlying attitudes, values and behaviour within and between European nations. However, recent empirical evidence has shown that there is a generally decreasing trend in response rates over the rounds in ESS \citep{Stoop2010, koen2018response}. The declining response rates of ESS may inflate its administration costs, degrade response quality and cast doubt on the validity of inferences. To meet these challenges, modern survey researchers have utilized SQD to collect survey data, which allows for different subsets of questionnaire to be collected from different respondents. In this section, the proposed method is applied to a SQD-type data created from the complete ESS dataset of round 8. For illustrating the application of the proposed approach, we focus on the state of health of democracies and its relevant factors with a shorter version of the real ESS questionnaire.

\begin{table}[!h]
  \centering
  \caption{Details of the covariates.}
  \resizebox{\textwidth}{!}{
    \begin{tabular}{lll}
    \hline
    Module & Variable  & Definition \\
    \hline
    A     & $X_1$  & Years of full-time education completed \\
         & $X_2$  & Age of respondent \\
         & $X_3$  & Gender \\
         \hline
    B     & $X_4$  & Trust in the legal system \\
         & $X_5$  & Trust in the police \\
         & $X_6$  & Trust in country's parliament \\
         & $X_7$  & Trust in political parties \\
         & $X_8$  & Trust in the United Nations \\
         & $X_9$  & Trust in the European Parliament \\
         & $X_{10}$ & Trust in politicians \\
         & $X_{11}$ & How satisfied with present state of economy in country \\
         & $X_{12}$ & State of education in country nowadays \\
         & $X_{13}$ & How satisfied with the national government \\
         & $X_{14}$ & Political system allows people to have a say in what government does \\
         & $X_{15}$ & Political system allows people to have influence on politics \\
         & $X_{16}$ & Political system in country ensures everyone fair chance to participate in politics \\
         & $X_{17}$ & Government in country takes into account the interests of all citizens \\
         & $X_{18}$ & Decisions in country politics are transparent \\
         \hline
    C     & $X_{19}$ & Approve if person chooses never to have children  \\
         & $X_{20}$ & Approve if person lives with partner not married to  \\
         & $X_{21}$ & Approve if person have child with partner not married to  \\
         & $X_{22}$ & Approve if person has full-time job while children aged under 3  \\
         & $X_{23}$ & Approve if person gets divorced while children aged under 12  \\
         \hline
    D     & $X_{24}$ & Household's total net income, all sources \\
         & $X_{25}$ & Compared other people in country, fair chance get job I seek \\
         & $X_{26}$ & How happy are you \\
         & $X_{27}$ & Influence decision to recruit in country: person's knowledge and skills \\
         & $X_{28}$ & Compared other people in country, fair chance achieve level of education I seek \\
         & $X_{29}$ & Feeling about household's income nowadays \\
         \hline
    \end{tabular}%
    }
  \label{tab:real:variable}%
\end{table}%

In our analysis, the response variable is satisfaction with democracy (SWD), which is measured by 11-point Likert scale from 0 to 10. As in some social science research \citep{Moore:2006, Wu:2017}, we also consider a linear regression model for predicting SWD but with SQD-type data. There are 29 covariates incorporated into the model, and we divide these covariates into four modules labeled A to D, which consist of covariates related to different aspects of survey participants. For example, module A contains three basic variables, that are age, gender and education status, and module B contains 15 covariates related to the participants' attitude to politics and society. Table~\ref{tab:real:variable} presents the detailed definitions of these 29 covariates. After excluding the samples with non-response record and incomprehensible results, we finally obtain a complete dataset containing 27341 valid cases.

\begin{figure}[!htbp]
\centering
\vspace{-0.35cm}	
\subfigtopskip=2pt
\subfigbottomskip=2pt
\subfigcapskip=-5pt
\setlength{\abovecaptionskip}{-2pt}
\subfigure[]{
\includegraphics[width=0.45\linewidth]{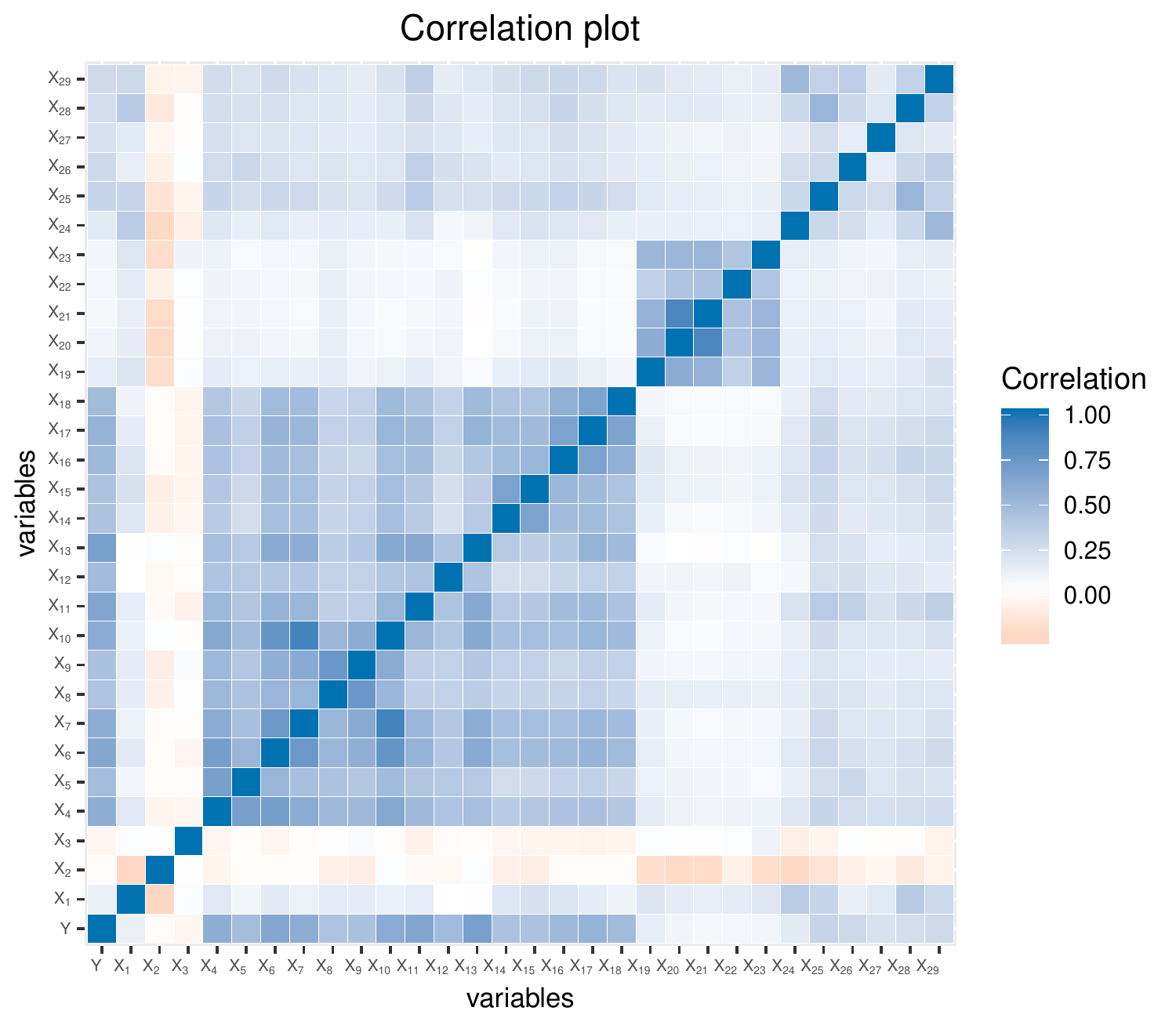}}\quad
\subfigure[]{
\includegraphics[width=0.45\linewidth]{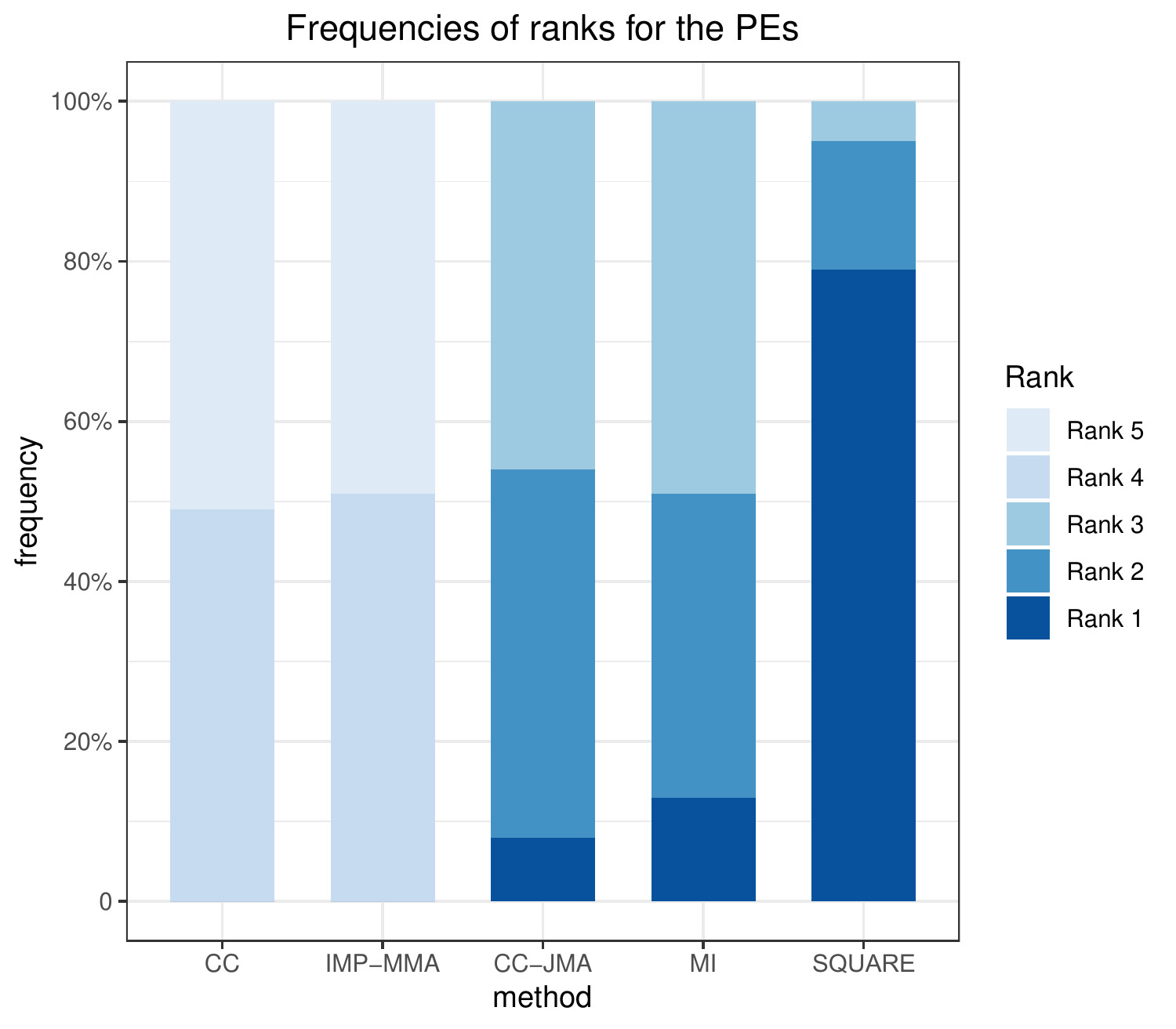}}	

\caption{Data analysis. (a) Correlation plot based on the complete dataset. (b) The relative frequencies of ranks for the PEs of competing methods over 100 replications.}
\label{fig:real}
\end{figure}

From Figure~\ref{fig:real} (a), we can see that the modular structure also corresponds to an obvious correlation structure between covariates: the covariates in the same module are more correlated than the covariates in different modules. Actually, the means of the absolute value of inner-modules correlations and between-modules correlations are 0.40 and 0.13 respectively. In our SQD-type, the covariates in module A are assigned into the common set. The remaining 26 questions are split into 3 subsets according to the modular structure. Then three different sub-questionnaires are formed by combining the common set and one of the other three sets. The sample sizes for the complete questionnaire and the sub-questionnaires are set to be $n_0 = 50$ and $n_1=n_2= n_3=1000$ respectively. The participants are randomly selected and assigned to one of the four questionnaires. The cases left are used as testing data to evaluate the prediction accuracy of the considered methods. This process is repeated
100 times.

The average prediction errors (PE) over 100 replications are computed for different methods, which are CC: 7.048 (1.881), MI: 3.093 (0.114), IMP-MMA: 6.531 (0.172), CC-JMA: 3.206 (0.466) and SQUARE: 2.886 (0.211), respectively. For comparison, we calculate the full data estimate based on 3050 randomly selected cases without missing, and its PE is 2.702 (0.012). It is observed that SQUARE performs much  better than the alternatives using SQD-type data,  but comparable to the full data estimate. Furthermore, we rank the PEs of the five competing methods at each replication, the results are presented in Figure ~\ref{fig:real} (b). Among the 100 replications, our proposed method ranks $\#$1 in 78\% replications and ranks top 2 in 96\% replications, which means that in most replications, the proposed method yields the smallest PE among all competitors. This reconfirms the superiority of the proposed method in SQD-type data analysis.

\begin{figure}[!htbp]
\centering
\includegraphics [angle=-0, scale=0.65]{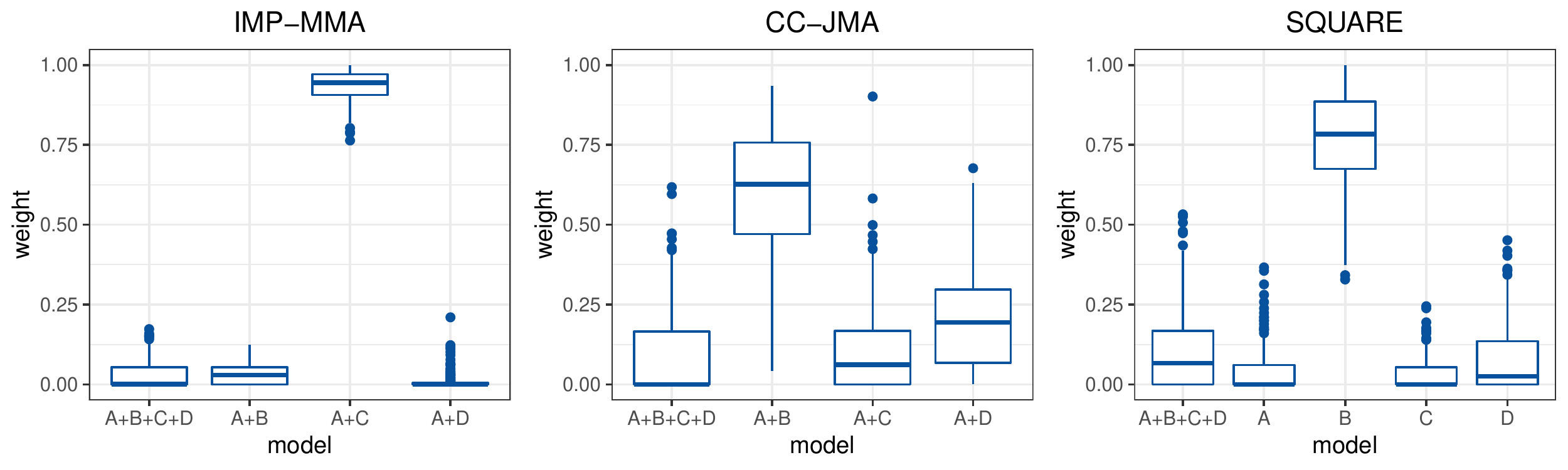}\par
\caption{Results from the analysis of ESS dataset. Estimated weights of candidate models for three model averaging methods.}
\label{fig:weight}
\end{figure}

For the methods of model averaging, the estimated weights are presented in Figure~\ref{fig:weight}.
There are five candidate models combined in SQUARE, and the candidate model with module B is assigned the largest weight. It implies that covariates in module B are most predictive to the response variable SWD, which is basically in accordance with the correlation results presented in  Figure~\ref{fig:real} (a). Meanwhile, the items in module B are related to the social and political trust of respondents and ought to be more informative in predicting SWD. In addition, IMP-MMA and CC-JMA fit the same four candidate models, but assign different weights. Similar to SQUARE, CC-JMA assigns the candidate with modules A and B the largest weight, while IMP-MMA puts most of weight on the candidate based on modules A and C. The bias of IMP-MMA, caused by replacing the missing values with zeros, is one of the contributors to arriving at different conclusions and leading to worse prediction performance than SQUARE and CC-JMA. Overall, the proposed method seems more recommendable for SQD-type data analysis.

\section{Discussion}\label{sec:discussion}
SQD is a relatively new survey tool to reduce survey burden, decrease nonresponse rates and improve quality of survey data. Data of similar block structure have also been increasingly seen in various applications. There is a natural demand for statistical methods capable of analyzing the SQD-type data. In this article, we have proposed a novel approach (SQUARE) to estimate the regression function, which works by constructing a list of candidate models based on data-blocks and combining them with different weights according to their relative contributions on the complete cases. Different from the studies of missing data, the proposed method avoids using imputation to fill out the unobserved data, and hence is more stable and computationally efficient. Meanwhile, it also differs significantly from the least square averaging estimators of missing data \citep{Zhang:2013,Fang:2017} by allowing for a large class of linear estimators to capture the characteristics of the true regression function with precision. Another important feature is that SQUARE allows the model wights to vary freely between 0 and 1 without the standard constraint of summing up to 1,
so that the strengths of candidate models can be efficiently combined and the biases can be largely canceled out. The results of asymptotical optimality have provided a strong basis for the adaptive strengths and superiority over the competing methods.  Numerical studies have shown that the proposed approach has better performance in our simulations and real data analysis as well.

There are several directions for further research. The focus of this article has been on a simple SQD-type structure, in which each partial set of covariates can be observed only for a distinct sub-sample of the data. A further research interest would be to estimate the regression function based on more general SQD-type data structures, in which observed data-blocks have more overlaps in samples and variables. Moreover, this article focuses primarily on the estimation of the regression function, and we have implicitly assumed the number of covariates is fixed and smaller than $n_0$. A natural question is how to select the important variables in regression based on such SQD-type data. Another extension, particularly with quite different targets from those in regression, is to extend the
analysis to clustering, where the groups of individuals with distinct characteristics can be found. We leave these topics for future research.

\section{Acknowledgements}
Li's work is supported by the National Natural Science Foundation of China (71771211) and National Bureau of Statistics of China (2019LD07). Lin's work is supported by National Statistical Science Research Project(2019LZ22) and foundation from Ministry of Education of China (No.20JZD023).

\bigskip
\begin{center}
{\large\bf Appendix A. Proof of Theorem \ref{theo:asyopt}\label{appendix}}
\end{center}

For notational convenience, we define $\mathcal{N}_0=S_0, \mathcal{N}_1=\bigcup_{m=1}^MS_m$ and $\mathcal{N}_{m+1}=S_m$ for $m\geq 1$. We sometimes will also liberally share the bounding constants ($C$, $C'$, etc.) when deriving inequalities. Let $\bH_m=\tilde{\bH}_m$ for $m\geq 1$. Define $\tilde{L}_n(\bw)=\|\tilde{\bmu}_{\mathcal{N}_0}(\bw)-\bmu_{\mathcal{N}_0}\|^2=\|\sum_{m=0}^{M+1}w_m\tilde{\bH}_m\by_{\mathcal{N}_m}-\bmu_{\mathcal{N}_0}\|^2$. Note that the criterion in (\ref{eq:criterion}) can be written as
\begin{equation*}\label{eq:decomp_cn}
  \|\bepsi_{\mathcal{N}_0}\|^2+L_n(\bw)\left(\frac{\tilde{L}_n(\bw)}{L_n(\bw)}+\frac{2\langle \bepsi_{\mathcal{N}_0}, \bmu_{\mathcal{N}_0}-\tilde{\bmu}_{\mathcal{N}_0}(\bw)\rangle/R_n(\bw)}{L_n(\bw)/R_n(\bw)}\right).
\end{equation*}
The proof of the asymptotic optimality of $\hat{\bw}$ falls naturally into three parts:
\begin{equation}\label{eq:goal2}
  \sup_{\bw\in Q_n}\left|\tilde{L}_n(\bw)/L_n(\bw)-1\right|\stackrel{p}{\to} 0,
\end{equation}
\begin{equation}\label{eq:goal3}
  \sup_{\bw \in Q_n}\left|\langle\bepsi_{\mathcal{N}_0}, \bmu_{\mathcal{N}_0}-\tilde{\bmu}_{\mathcal{N}_0}(\bw)\rangle\right|/R_n(\bw)\stackrel{p}{\to} 0,
\end{equation}
\begin{equation}\label{eq:goal1}
  \sup_{\bw\in Q_n}\left|L_n(\bw)/R_n(\bw)-1\right|\stackrel{p}{\to} 0.
\end{equation}

Our first goal is to show (\ref{eq:goal2}). Note that
\begin{equation*}
\begin{split}
    \tilde{L}_n(\bw)-L_n(\bw)=&\|w_0\tilde{\bH}_0\by_{\mathcal{N}_0}\|^2-\|w_0\bH_0\by_{\mathcal{N}_0}\|^2+2\langle w_0\tilde{\bH}_0\by_{\mathcal{N}_0},\sum_{m=1}^{M+1}w_m\tilde{\bH}_m\by_{\mathcal{N}_m}-\bmu_{\mathcal{N}_0}\rangle\\
    &-2\langle w_0\bH_0\by_{\mathcal{N}_0},\sum_{m=1}^{M+1}w_m\bH_m\by_{\mathcal{N}_m}-\bmu_{\mathcal{N}_0}\rangle\\
    =&w_0^2\|(\tilde{\bH}_0-\bH_0)\by_{\mathcal{N}_0}\|^2+2w_0\langle(\tilde{bH}_0-\bH_0)\by_{\mathcal{N}_0},\sum_{m=0}^{M+1}w_m\bH_m\by_{\mathcal{N}_m}-\bmu_{\mathcal{N}_0}\rangle\\
    \leq &w_0^2\|(\tilde{\bH}_0-\bH_0)\by_{\mathcal{N}_0}\|^2+2w_0\|(\tilde{\bH}_0-\bH_0)\by_{\mathcal{N}_0}\|\sqrt{L_n\left(\bw\right)}
\end{split}
\end{equation*}
Therefore, it remains to show that
\begin{equation*}\label{eq:goal2suf}
\sup_{\bw\in Q_n}\frac{\|(\tilde{\bH}_0-\bH_0)\by_{\mathcal{N}_0}\|^2}{L_n(\bw)}\stackrel{p}{\to} 0.
\end{equation*}
And note that
\begin{equation*}
  \|(\tilde{\bH}_0-\bH_0)\by_{\mathcal{N}_0}\|^2=\|(\tilde{\bH}_0-\bH_0)\bmu_{\mathcal{N}_0}+(\tilde{\bH}_0-\bH_0)\bepsi_{\mathcal{N}_0}\|^2\leq 2\|(\tilde{\bH}_0-\bH_0)\bmu_{\mathcal{N}_0}\|^2+2\|(\tilde{\bH}_0-\bH_0)\bepsi_{\mathcal{N}_0}\|^2.
\end{equation*}
It remains to prove that
\begin{equation}\label{eq:goal2suf1}
  \frac{\|(\tilde{\bH}_0-\bH_0)\bmu_{\mathcal{N}_0}\|^2}{\zeta_n}\to 0,
\end{equation}
\begin{equation}\label{eq:goal2suf2}
  \frac{\|(\tilde{\bH}_0-\bH_0)\bepsi_{\mathcal{N}_0}\|^2}{\zeta_n}\to_p 0.
\end{equation}
We have
\begin{equation*}
    \frac{\|(\tilde{\bH}_0-\bH_0)\bmu_{\mathcal{N}_0}\|^2}{\zeta_n}\leq
    \frac{\left[\lambda_{\max}(\tilde{\bH}_0-\bH_0)\right]^2\|\bmu_{\mathcal{N}_0}\|^2}{\zeta_n}\leq
    \frac{Cp^2}{n_0\zeta_{n}}\to 0.
\end{equation*}
Thus, (\ref{eq:goal2suf1}) is proved. Let $\bSigma_m$ denote $E\bepsi_{\mathcal{N}_m}\bepsi_{\mathcal{N}_m}^T$ for $0\leq m \leq M+1$. Then, for any $\delta>0$, with Markov inequality, we have
\begin{equation*}
\begin{split}
    P&\left(\frac{\|\tilde{\bH}_0-\bH_0)\bepsi_{\mathcal{N}_0}\|^2}{\zeta_n}>\delta\right)\leq \frac{E\|(\tilde{H}_0-\bH_0)\bepsi_{\mathcal{N}_0}\|^2}{\delta\zeta_n}\\
    &=\frac{{\rm tr}\left[(\tilde{\bH}_0-\bH_0)\bSigma_0(\tilde{\bH}_0-\bH_0)^T\right]}{\delta\zeta_n}\leq \frac{Cp^2}{n_0\delta\zeta_n}\to 0,\\
\end{split}
\end{equation*}
which proves (\ref{eq:goal2suf2}). Thus, (\ref{eq:goal2}) is proved.

Our second goal is to show (\ref{eq:goal3}). Because
\begin{equation*}
\begin{split}
    &|\langle\bepsi_{\mathcal{N}_0}, \bmu_{\mathcal{N}_0}-\tilde{\bmu}_{\mathcal{N}_0}(\bw)\rangle|=|\langle\bepsi_{\mathcal{N}_0}, \bmu_{\mathcal{N}_0}-\sum_{m=0}^{M+1}w_m\tilde{\bH}_m\bmu_{\mathcal{N}_m}-\sum_{m=0}^{M+1}w_m\tilde{\bH}_m\bepsi_{\mathcal{N}_m}\rangle|\\
    &\leq |\langle\bepsi_{\mathcal{N}_0},\bmu_{\mathcal{N}_0}\rangle|+|\langle\bepsi_{\mathcal{N}_0},\sum_{m=0}^{M+1}w_m\tilde{\bH}_m\mu_{\mathcal{N}_m}\rangle|+|\langle\bepsi_{\mathcal{N}_0},\sum_{m=0}^{M+1}w_m\tilde{\bH}_m\bepsi_{\mathcal{N}_m}\rangle|,
\end{split}
\end{equation*}
the proof of (\ref{eq:goal3}) can be divided into three steps:
\begin{equation}\label{eq:goal3suf1}
  \frac{|\langle\bepsi_{\mathcal{N}_0},\bmu_{\mathcal{N}_0}\rangle|}{\zeta_n}\to_p 0,
\end{equation}
\begin{equation}\label{eq:goal3suf2}
  \sup_{\bw\in Q_n}\frac{|\langle\bepsi_{\mathcal{N}_0},\sum_{m=0}^{M+1}w_m\tilde{\bH}_m\bmu_{\mathcal{N}_m}\rangle|}{\zeta_n}\to_p 0,
\end{equation}
\begin{equation}\label{eq:goal3suf3}
    \sup_{\bw\in Q_n}\frac{|\langle\bepsi_{\mathcal{N}_0},\sum_{m=0}^{M+1}w_m\tilde{\bH}_m\bepsi_{\mathcal{N}_m}\rangle|}{\zeta_n}\to_p 0.
\end{equation}
For any $\delta>0$,
\begin{equation*}
\begin{split}
    P&\left(\frac{|\langle\bepsi_{\mathcal{N}_0},\bmu_{\mathcal{N}_0}\rangle|}{\zeta_n}>\delta\right)
    \leq\frac{E\langle\bepsi_{\mathcal{N}_0},\bmu_{\mathcal{N}_0}\rangle^2}{\zeta_n^2\delta^2}\leq \frac{\lambda_{\max}(\bSigma_0)\|\bmu_{\mathcal{N}_0}\|^2}{\zeta_n^2\delta^2}\leq \frac{Cn_0}{\zeta_n^2\delta^2}\to 0,
\end{split}
\end{equation*}
which proves (\ref{eq:goal3suf1}). We can bound
\begin{equation}\label{eq:goal3suf2suf1}
  \begin{split}
    &|\langle\bepsi_{\mathcal{N}_0},\sum_{m=0}^{M+1}w_m\tilde{\bH}_m\bmu_{\mathcal{N}_m}\rangle|\leq \sum_{m=0}^{M+1}|\langle\bepsi_{\mathcal{N}_0},w_m\tilde{\bH}_m\bmu_{\mathcal{N}_m}\rangle|\\
    &\leq |\langle\bepsi_{\mathcal{N}_0},w_0\tilde{\bH}_0\mu_{\mathcal{N}_0}\rangle|+(M+1)\max_{1\leq m\leq M+1}|\langle\bepsi_{\mathcal{N}_0},\tilde{\bH}_m\bmu_{\mathcal{N}_m}\rangle|.
  \end{split}
\end{equation}
Note that for any $\delta>0$,
\begin{equation}\label{eq:goal3suf2suf2}
  \begin{split}
    P&\left(\frac{|\langle\bepsi_{\mathcal{N}_0},w_0\tilde{\bH}_0\bmu_{\mathcal{N}_0}\rangle|}{\zeta_n}>\delta\right)\leq \frac{E\langle\bepsi_{\mathcal{N}_0},w_0\tilde{\bH}_0\bmu_{\mathcal{N}_0}\rangle^2}{\delta^2\zeta_n^2}=\frac{\bmu_{\mathcal{N}_0}^T\tilde{\bH}_0^T\bSigma_0\tilde{\bH}_0\bmu_{\mathcal{N}_0}}{\delta^2\zeta_n^2}\leq \frac{C_1\bmu_{\mathcal{N}_0}^T\tilde{\bH}_0^T\tilde{\bH}_0\bmu_{\mathcal{N}_0}}{\delta^2\zeta_n^2}\\
    &\leq \frac{C_1\left[\lambda_{\max}(\tilde{\bH}_0)\right]^2\|\bmu_{\mathcal{N}_0}\|^2}{\delta^2\zeta_n^2}\leq \frac{C_1(1+C_2p/n_0)^2n_0}{\delta^2\zeta_n^2}
    \asymp \frac{n_0}{\zeta_n^2}+\frac{p}{\zeta_n^2}+\frac{p^2}{n_0\zeta_n^2}\to 0,
  \end{split}
\end{equation}
and
\begin{equation}\label{eq:goal3suf2suf3}
  \begin{split}
    &P\left[\frac{(M+1)\max_{1\leq m\leq M+1}|\langle\bepsi_{\mathcal{N}_0},\tilde{\bH}_m\bmu_{\mathcal{N}_m}\rangle|}{\zeta_n}>\delta\right]\leq \sum_{m=1}^{M+1}P\left[\frac{(M+1)|\langle\bepsi_{\mathcal{N}_0},\tilde{\bH}_m\bmu_{\mathcal{N}_m}\rangle|}{\zeta_n}>\delta \right]\\
    &\leq\frac{(M+1)^2}{\zeta_n^2\delta^2}\sum_{m=1}^{M+1}E\langle\bepsi_{\mathcal{N}_0},\tilde{\bH}_m\bmu_{\mathcal{N}_m}\rangle^2\leq \frac{(M+1)^2}{\zeta_n^2\delta^2}\sum_{m=1}^{M+1}C\left[\lambda_{\max}(\tilde{\bH}_m^T\tilde{\bH}_m)\right]^2\|\bmu_{\mathcal{N}_m}\|^2\\
    &\leq \frac{C(M+1)^3\bar{n}_M}{\zeta_n^2\delta^2}\to 0.
  \end{split}
\end{equation}
Combining (\ref{eq:goal3suf2suf2}) and (\ref{eq:goal3suf2suf3}) with (\ref{eq:goal3suf2suf1}), (\ref{eq:goal3suf2}) is proved. Similarly, we can bound
\begin{equation}\label{eq:goal3suf3suf1}
  \begin{split}
    &|\langle\bepsi_{\mathcal{N}_0},\sum_{m=0}^{M+1}w_m\tilde{\bH}_m\bepsi_{\mathcal{N}_m}\rangle|\leq |\langle\bepsi_{\mathcal{N}_0},w_0\tilde{\bH}_0\bepsi_{\mathcal{N}_0}\rangle|+\sum_{m=1}^{M+1}|\langle\bepsi_{\mathcal{N}_0},w_m\tilde{\bH}_m\bepsi_{\mathcal{N}_m}\rangle|\\
    &\leq |\langle\bepsi_{\mathcal{N}_0},w_0\tilde{\bH}_0\bepsi_{\mathcal{N}_0}\rangle|+(M+1)\max_{1\leq m\leq M+1}|\langle\bepsi_{\mathcal{N}_0},\tilde{\bH}_m\bepsi_{\mathcal{N}_m}\rangle|.
  \end{split}
\end{equation}
Note that for any $\delta>0$,
\begin{equation}\label{eq:goal3suf3suf2}
\begin{split}
    P\left(\frac{|\langle\bepsi_{\mathcal{N}_0},\tilde{\bH}_0\bepsi_{\mathcal{N}_0}\rangle|}{\zeta_n}>\delta\right)
    \leq \frac{E\langle\bepsi_{\mathcal{N}_0},\tilde{\bH}_0\bepsi_{\mathcal{N}_0}\rangle^2}{\zeta_n^2\delta^2}\leq \frac{{C\rm tr}(\tilde{\bH}_0\tilde{\bH}_0^T)}{\zeta_n^2\delta^2}\leq \frac{Cp}{\zeta_n^2\delta^2}\to 0,
\end{split}
\end{equation}
and
\begin{equation}\label{eq:goal3suf3suf3}
  \begin{split}
    &P\left[\frac{(M+1)\max_{1\leq m\leq M+1}|\langle\bepsi_{\mathcal{N}_0},\tilde{\bH}_m\bepsi_{\mathcal{N}_m}\rangle|}{\zeta_n}>\delta\right] \leq \sum_{m=1}^{M+1}P\left(\frac{(M+1)|\langle\bepsi_{\mathcal{N}_0},\tilde{\bH}_m\bepsi_{\mathcal{N}_m}\rangle|}{\zeta_n}>\delta\right)\\
    &\leq \frac{(M+1)^2}{\zeta_n^2\delta^2}\sum_{m=1}^{M+1}E\langle\bepsi_{\mathcal{N}_0},\tilde{\bH}_m\bepsi_{\mathcal{N}_m}\rangle^2
    \leq \frac{(M+1)^2}{\zeta_n^2\delta^2}\sum_{m=1}^{M+1}\lambda_{\max}(\bSigma_m)E\bepsi_{\mathcal{N}_0}^T\tilde{\bH}_m\tilde{\bH}_m^T\bepsi_{\mathcal{N}_0}\\
    &\leq \frac{(M+1)^2}{\zeta_n^2\delta^2}\sum_{m=1}^{M+1}\lambda_{\max}(\bSigma_m)\lambda_{\max}(\bSigma_0){\rm tr}(\tilde{\bH}_m\tilde{\bH}_m^T)
    \leq\frac{(M+1)^2}{\zeta_n^2\delta^2}\sum_{m=1}^{M+1}Cp_m\leq \frac{C(M+1)^2p}{\zeta_n^2\delta^2}\to 0.
  \end{split}
\end{equation}
Combining (\ref{eq:goal3suf3suf2}) and (\ref{eq:goal3suf3suf3}) with (\ref{eq:goal3suf3suf1}), (\ref{eq:goal3suf3}) is proved. Thus, the proof of (\ref{eq:goal3}) is completed.

Lastly, we show (\ref{eq:goal1}) holds. By a simple calculation, we have
\begin{equation*}
\begin{split}
    L_n(\bw)-R_n(\bw)=&\|\sum_{m=0}^{M+1}w_m\bH_m\bepsi_{\mathcal{N}_m}\|^2-E\|\sum_{m=0}^{M+1}w_m\bH_m\bepsi_{\mathcal{N}_m}\|^2\\
    &+2\langle\sum_{m=0}^{M+1}w_m\bH_m\bmu_{\mathcal{N}_m}-\bmu_{\mathcal{N}_0},\sum_{m=0}^{M+1}w_m\bH_m\bepsi_{\mathcal{N}_m}\rangle.
\end{split}
\end{equation*}
To prove~(\ref{eq:goal1}), it is sufficient to show
\begin{equation}\label{eq:goal1suf1}
  \sup_{\bw\in Q_n}\left|\|\sum_{m=0}^{M+1}w_m\bH_m\bepsi_{\mathcal{N}_m}\|^2-E\|\sum_{m=0}^{M+1}w_m\bH_m\bepsi_{\mathcal{N}_m}\|^2\right|/R_n(\bw)\stackrel{p}{\to} 0,
\end{equation}
and
\begin{equation}\label{eq:goal1suf2}
  \sup_{\bw\in Q_n}\left|\langle\sum_{m=0}^{M+1}w_m\bH_m\bmu_{\mathcal{N}_m}-\bmu_{\mathcal{N}_0},\sum_{m=0}^{M+1}w_m\bH_m\bepsi_{\mathcal{N}_m}\rangle\right|/R_n(\bw)\stackrel{p}{\to} 0.
\end{equation}
We first establish claim (\ref{eq:goal1suf1}). Using the triangle inequality and the independence between $\bepsi_{\mathcal{N}_0}$ and $\bepsi_{\mathcal{N}_m}$ for $ m \geq 1$, we can bound the numerator by
\begin{equation*}
  \begin{split}
    &\left|\|\sum_{m=0}^{M+1}w_m\bH_m\bepsi_{\mathcal{N}_m}\|^2-E\|\sum_{m=0}^{M+1}w_m\bH_m\bepsi_{\mathcal{N}_m}\|^2\right|\leq \left|\|w_0\bH_0\bepsi_{\mathcal{N}_0}\|^2-E\|w_0\bH_0\bepsi_{\mathcal{N}_0}\|^2\right|\\
    &+\left|2\langle w_0\bH_0\bepsi_{\mathcal{N}_0},\sum_{m=1}^{M+1}w_m\bH_m\bepsi_{\mathcal{N}_m}\rangle\right|+\left|\|\sum_{m=1}^{M+1}w_m\bH_m\bepsi_{\mathcal{N}_m}\|^2-E\|\sum_{m=1}^{M+1}w_m\bH_m\bepsi_{\mathcal{N}_m}\|^2\right|.
  \end{split}
\end{equation*}
Then, the proof of (\ref{eq:goal1suf1}) will be divided into three parts:
\begin{equation}\label{eq:goal1suf1suf1}
  \sup_{\bw\in Q_n}\frac{\left|\|\bH_0\bepsi_{\mathcal{N}_0}\|^2-E\|\bH_0\bepsi_{\mathcal{N}_0}\|^2\right|}{R_n(\bw)}\stackrel{p}{\to} 0,
\end{equation}
\begin{equation}\label{eq:goal1suf1suf2}
    \sup_{\bw\in Q_n}\frac{\left|\langle w_0\bH_0\bepsi_{\mathcal{N}_0},\sum_{m=1}^{M+1}w_m\bH_m\bepsi_{\mathcal{N}_m}\rangle\right|}{R_n(\bw)}\stackrel{p}{\to} 0,
\end{equation}
\begin{equation}\label{eq:goal1suf1suf3}
    \sup_{\bw\in Q_n}\frac{\left|\|\sum_{m=1}^{M+1}w_m\bH_m\bepsi_{\mathcal{N}_m}\|^2-E\|\sum_{m=1}^{M+1}w_m\bH_m\bepsi_{\mathcal{N}_m}\|^2\right|}{R_n(\bw)}\stackrel{p}{\to} 0.
\end{equation}
We first prove (\ref{eq:goal1suf1suf1}). Using Chebyshev's inequality and \textbf{(C4)} in Theorem~1, we observe, for any $\delta>0$, that
\begin{equation}\nonumber
\begin{split}
      P&\left(\sup_{\bw\in Q_n}\frac{\left|\|\bH_0\bepsi_{\mathcal{N}_0}\|^2-E\|\bH_0\bepsi_{\mathcal{N}_0}\|^2\right|}{R_n(\bw)}>\delta\right)\leq P\left(\frac{\left|\|\bH_0\bepsi_{\mathcal{N}_0}\|^2-E\|\bH_0\bepsi_{\mathcal{N}_0}\|^2\right|}{\zeta_n}>\delta\right)\\
      &\leq \frac{E\left(\|\bH_0\bepsi_{\mathcal{N}_0}\|^2-E\|\bH_0\bepsi_{\mathcal{N}_0}\|^2\right)^2}{\delta^2\zeta_n^2}
      \leq \frac{E\|\bH_0\bepsi_{\mathcal{N}_0}\|^4}{\delta^2\zeta_n^2}\leq \frac{C\mbox{tr}(\bH_0)}{\delta^2\zeta_n^2}\leq\frac{Cp}{\delta^2\zeta_n^2}\to 0,
\end{split}
\end{equation}
where $C$ is some constant. When we establish (\ref{eq:goal1suf1suf2}), it suffices to show
\begin{equation}\nonumber
  \begin{split}
    P&\left(\frac{(M+1)\max_{1\leq m\leq M+1}|\langle \bH_0\bepsi_{\mathcal{N}_0},\bH_m\bepsi_{\mathcal{N}_m}\rangle|}{\zeta_n}>\delta \right)\leq \sum_{m=1}^{M+1}P\left(\langle \bH_0\bepsi_{\mathcal{N}_0},\bH_m\bepsi_{\mathcal{N}_m}\rangle^2>\frac{\delta^2\zeta_n^2}{(M+1)^2}\right)\\
    &\leq \frac{(M+1)^2}{\delta^2\zeta_n^2}\sum_{m=1}^{M+1}E\langle \bH_0\bepsi_{\mathcal{N}_0},\bH_m\bepsi_{\mathcal{N}_m}\rangle\leq \frac{C(M+1)^2}{\delta^2\zeta_n^2}\sum_{m=1}^{M+1}\mbox{tr}(\bH_m^T\bH_m)\leq\frac{C(M+1)^2p}{\delta^2\zeta_n^2}\to 0.
  \end{split}
\end{equation}
To establish (\ref{eq:goal1suf1suf3}), we can bound the numerator by
\begin{equation}\nonumber
  \begin{split}
    &\left|\|\sum_{m=1}^{M+1}w_m\bH_m\bepsi_{\mathcal{N}_m}\|^2-E\|\sum_{m=1}^{M+1}w_m\bH_m\bepsi_{\mathcal{N}_m}\|^2\right|\leq \left|\sum_{m=1}^{M+1}w_m^2\|\bH_m\bepsi_{\mathcal{N}_m}\|^2-E\sum_{m=1}^{M+1}w_m^2\|\bH_m\bepsi_{\mathcal{N}_m}\|^2\right|\\
    &+2\left|\sum_{m=1}^{M}\sum_{k=m+1}^{M+1}w_mw_k\langle \bH_m\bepsi_{\mathcal{N}_m},\bH_k\bepsi_{\mathcal{N}_k}\rangle-E\sum_{m=2}^{M+1}w_1w_{m}\langle \bH_1\bepsi_{\mathcal{N}_1},\bH_m\bepsi_{\mathcal{N}_{m}}\rangle\right|
  \end{split}
\end{equation}
Using the triangle inequality, Bonferroni's inequality, Chebyshev's inequality, we observe, for any $\delta>0$, that
\begin{equation*}
  \begin{split}
    P&\left(\sup_{\bw\in Q_n}\frac{\left|\sum_{m=1}^{M+1}w_m^2\|\bH_m\bepsi_{\mathcal{N}_m}\|^2-E\sum_{m=1}^{M+1}w_m^2\|\bH_m\bepsi_{\mathcal{N}_m}\|^2\right|}{\zeta_n}>\delta\right)\\
    &\leq P\left((M+1)\max_{1\leq m\leq M+1}\left|\|\bH_m\bepsi_{\mathcal{N}_m}\|^2-E\|\bH_m\bepsi_{\mathcal{N}_m}\|^2\right|>\delta\zeta_n\right)\\
    &\leq \frac{(M+1)^2}{\delta^2\zeta_n^2}\sum_{m=1}^{M+1}E\|\bH_m\bepsi_{\mathcal{N}_m}\|^4
    \leq \frac{C(M+1)^2}{\delta^2\zeta_n^2}\sum_{m=1}^{M+1}\lambda_{\max}(\bH_m\bH_m^T)\mbox{tr}(\bH_m^T\bH_m)\leq\frac{CM^2p}{\delta^2\zeta_n^2}\to 0.
  \end{split}
\end{equation*}
For any $\delta>0$, we observe that
\begin{equation*}
  \begin{split}
    P&\left(\sup_{\bw\in Q_n}\frac{\left|\sum_{m=1}^{M}\sum_{k=m+1}^{M+1}w_mw_k\langle \bH_m\bepsi_{\mathcal{N}_m},\bH_k\bepsi_{\mathcal{N}_k}\rangle\right|}{\zeta_n}>\delta\right)\\
    &\leq P\left(\frac{M(M+1)}{2}\max_{m\neq k}\left|\langle \bH_m\bepsi_{\mathcal{N}_m},\bH_k\bepsi_{\mathcal{N}_k}\rangle\right|>\delta\zeta_n\right)\leq\frac{M^2(M+1)^2}{4\delta^2\zeta_n^2}\sum_{m=1}^{M}\sum_{k=m+1}^{M+1}E\langle \bH_m\bepsi_{\mathcal{N}_m},\bH_k\bepsi_{\mathcal{N}_k}\rangle^2\\
    &\leq \frac{CM^2(M+1)^2}{4\delta^2\zeta_n^2}\sum_{m=1}^{M}\sum_{k=m+1}^{M+1}\lambda_{\max}(\bH_m\bH_m')\mbox{tr}(\bH_k'\bH_k)\leq \frac{CM^2(M+1)^3p}{4\delta^2\zeta_n^2}\to 0,
  \end{split}
\end{equation*}
and
\begin{equation*}
\begin{split}
    &P\left(    \sup_{\bw\in Q_n}\frac{\left|\sum_{m=2}^{M+1}w_1w_{m}\langle \bH_1\bepsi_{\mathcal{N}_1},\bH_{m}\bepsi_{\mathcal{N}_{m}}\rangle-E\sum_{m=2}^{M+1}w_1w_{m}\langle \bH_1\bepsi_{\mathcal{N}_1},\bH_{m}\bepsi_{\mathcal{N}_{m}}\rangle\right|}{\zeta_n}>\delta\right)\\
    &\leq P\left(M\max_{2\leq m\leq M+1}\left|\langle \bH_1\bepsi_{\mathcal{N}_1},\bH_{m}\bepsi_{\mathcal{N}_{m}}\rangle-E\langle \bH_1\bepsi_{\mathcal{N}_1},\bH_{m}\bepsi_{\mathcal{N}_{m}}\rangle\right|>\delta\zeta_n\right)\\
    &\leq \frac{M^2}{\delta^2\zeta_n^2}\sum_{m=2}^{M+1}E\langle \bH_1\bepsi_{\mathcal{N}_1}, \bH_{m}\bepsi_{\mathcal{N}_{m}} \rangle^2\leq \frac{M^2}{\delta^2\zeta_n^2}\sum_{m=2}^{M+1}E\|\bH_1\bepsi_{\mathcal{N}_1}\|^2\|\bH_{m}\bepsi_{\mathcal{N}_{m}}\|^2\\
    &\leq \frac{M^2}{\delta^2\zeta_n^2}\sum_{m=2}^{M+1}(E\|\bH_1\bepsi_{\mathcal{N}_1}\|^4)^{1/2}E(\|\bH_{m}\bepsi_{\mathcal{N}_{m}}\|^4)^{1/2}\leq \frac{CM^3p}{\delta^2\zeta_n^2}\to 0.
\end{split}
\end{equation*}
Thus, the proof of (\ref{eq:goal1suf1suf3}) is completed. The proof of (\ref{eq:goal1suf2}) follows the same idea of (\ref{eq:goal1suf1}). Thus, we skip some similar material here.

\bibliographystyle{apalike}

\bibliography{Bibliography-MM-MC}
\end{document}